\documentclass[prb,preprint,showpacs,floatfix]{revtex4}
\usepackage{amsmath}
\usepackage{epsfig,amssymb,txfonts}
\usepackage[hang,raggedright]{subfigure}

\newcommand*{\abinit}[0]{density-functional theory}

\newcommand*{\et}[0]{\textit{et~al.}}
\newcommand*{\eqn}[1]{Eqn.~\protect\eqref{eqn:#1}}
\newcommand*{\fig}[1]{Figure~\protect\ref{fig:#1}}
\newcommand*{\tab}[1]{Table~\protect\ref{tab:#1}}
\newcommand*{\sect}[1]{Section~\protect\ref{sec:#1}}
\newcommand*{\term}[1]{\textit{#1}}
\newcommand*{\INT}[0]{Gradshteyn and Ryzhik\cite{IntegralTables}}

\newcommand*{\eps}[0]{\varepsilon}
\newcommand*{\x}[0]{\times}

\newcommand*{\inv}[1]{{#1}^{-1}}
\DeclareMathOperator{\Si}{Si} 

\newcommand*{\CART}[1]{\underline{#1}}  
\newcommand*{\FFT}[1]{\widetilde{#1}} 
\newcommand*{\GL}[1]{\CART{G}_{#1}^{\textrm{L}}}
\newcommand*{\GE}[1]{\CART{G}_{#1}^{\textrm{E}}}
\newcommand*{\Gdc}[1]{\CART{G}_{#1}^{\textrm{dc}}}

\newcommand*{\D}[1]{\CART{D}_{#1}}
\newcommand*{\GLk}[1]{\FFT{G}_{#1}^{\rm L}}
\newcommand*{\GEk}[1]{\FFT{G}_{#1}^{\rm E}}
\newcommand*{\Gdck}[1]{\FFT{G}_{#1}^{\textrm{dc}}}
\newcommand*{\Gsck}[1]{\FFT{G}_{#1}^{\textrm{sc}}}
\newcommand*{\Dk}[1]{\FFT{D}_{#1}}
\newcommand*{\Le}[0]{\FFT{\Lambda}^{(2)}}
\newcommand*{\Lq}[0]{\FFT{\Lambda}^{(4)}}
\newcommand*{\fc}[0]{f_\textrm{cut}}
\newcommand*{\delZ}[1]{\Delta^{(0)}[#1]}
\newcommand*{\del}[1]{\Delta^{(2)}[#1]}
\newcommand*{\eGF}[0]{\CART{\eps}^{\textrm{GF}}}
\newcommand*{\C}[0]{\mathbf{C}}
\newcommand*{\kv}[0]{\vec k}
\newcommand*{\Rv}[0]{\vec R}
\newcommand*{\xv}[0]{\vec x}
\newcommand*{\uv}[0]{\vec u}
\newcommand*{\fv}[0]{\vec f}
\newcommand*{\tv}[0]{\vec t}
\newcommand*{\nhat}[0]{\hat n_\perp}
\newcommand*{\nperp}[0]{\vec n_\perp}
\newcommand*{\mperp}[0]{\vec m_\perp}
\newcommand*{\kmax}[0]{k_\textrm{max}}
\newcommand*{\ident}[0]{\mathbf{1}}
\newcommand*{\zero}[0]{\mathbf{0}}
\newcommand*{\Lmax}[0]{L_\textrm{max}}
\newcommand*{\Nmax}[0]{N_\textrm{max}}
\newcommand*{\Int}[2]{\int_{#1}^{#2}\displaylimits\negthickspace}
\newcommand*{\intsphere}[0]{\iint_{4\pi}\displaylimits\negthickspace}
\newcommand*{\intcirc}[0]{\Int{0}{2\pi}}
\newcommand*{\intkmax}[0]{\Int{0}{\kmax}}
\newcommand*{\intx}[0]{\int_0^{x}\negthickspace}
\newcommand*{\intBZ}[0]{\iiint_{\textrm{BZ}}\displaylimits\negthickspace}
\newcommand*{\intBZtwo}[0]{\iint_{\textrm{BZ}}\displaylimits\negthickspace}
\newcommand*{\intBZone}[0]{\int_{\textrm{BZ}}\displaylimits\negthickspace}
\newcommand*{\kperp}[0]{\vec k_\perp}
\newcommand*{\kpara}[0]{\vec k_\|}
\newcommand*{\kplane}[0]{\vec k_{\textrm{plane}}}
\newcommand*{\Rperp}[0]{R_\perp}
\newcommand*{\Nkpt}[0]{N_{\textrm{kpt}}}

\newcommand*{\xdR}[0]{\hat x\cdot\hat R}
\newcommand*{\lelas}[0]{l_{\textrm{elas}}}

\newcommand*{\be}[0]{\begin{equation}}
\newcommand*{\ee}[0]{\end{equation}}
\newcommand*{\beu}[0]{\begin{equation*}}
\newcommand*{\eeu}[0]{\end{equation*}}
\newcommand*{\ba}[0]{\begin{array}}
\newcommand*{\ea}[0]{\end{array}}
\newcommand*{\bfig}[0]{\begin{figure}}
\newcommand*{\efig}[0]{\end{figure}}
\newcommand*{\bfigwide}[0]{\begin{figure*}}
\newcommand*{\efigwide}[0]{\end{figure*}}

\newcommand*{\entry}[1]{\multicolumn{1}{c}{\underbar{#1}}}

\newcommand*{\tabpara}[2]{\parbox[c]{#1}{\flushleft#2 \vskip2pt}}
\newcommand*{\tabparac}[2]{\parbox[c]{#1}{#2}}
\newcommand*{\barspace}[0]{\\[2pt]\hline\\[-12pt]}

\newlength{\wholefigwidth}
\newlength{\smallfigwidth}
\newlength{\halfsmallfigwidth}
\begin{document}

\title{Lattice Green function for extended defect calculations:
Computation and error estimation with long-range forces}

\author{D. R. Trinkle}
\altaffiliation{Department of Material Science and Engineering,
University of Illinois, Urbana-Champaign, 1304 W. Green Street,
Urbana, IL 61801, USA}
\email{dtrinkle@uiuc.edu}
\affiliation{Materials and Manufacturing Directorate, Air Force
Research Laboratory, Wright Patterson Air Force Base, Dayton, Ohio
45433-7817}

\date{\today}

\begin{abstract}
Computing the atomic geometry of lattice defects---point defects,
dislocations, crack tips, surfaces, or boundaries---requires an
accurate coupling of the local strain field to the long-range elastic
field.  Periodic boundary conditions used by classical potentials or
\abinit\ may not accurately reproduce the correct bulk response to an
isolated defect; this is especially true for dislocations.  Recently,
flexible boundary conditions have been developed to produce the
correct long-range strain field from a defect---effectively
``embedding'' a defect in a finite cell with infinite bulk response,
isolating it from either periodic images or free surfaces.  Flexible
boundary conditions require the calculation of the bulk response with
the lattice Green function (LGF).  While the LGF can be computed from
the dynamical matrix, for supercell methods (periodic boundary
conditions) it can only be calculated up to a maximum range.  We
illustrate how to accurately calculate the lattice Green function and
estimate the error using a cutoff dynamical matrix combined with
knowledge of the long-range behavior of the lattice Green function.
The effective range of deviation of the lattice Green function from
the long-range elastic behavior provides an important length scale in
multiscale quasi-continuum and flexible boundary-condition
calculations, and measures the error introduced with periodic boundary
conditions.
\end{abstract}


\maketitle

\section{Introduction}

Lattice defects---e.g., interstitials, vacancies, dislocations, crack
tips, free surfaces, interfaces, and boundaries---each play key roles
in material properties,\cite{Haasen96} and in order to understand
defects, one must know their geometry.  The far-field geometry for
many defects is accurately described by anisotropic elasticity
theory.\cite{Stroh1962,Bacon1979} However, the elastic solution often
diverges near the atomic-scale center of the defect, and in many cases
the center is difficult to investigate with current microscopy
techniques.  This is especially true of dislocations, which control
plasticity in metals\cite{Haasen96} and can severely limit device
utility in semiconductors.\cite{Rudolph1999} Only recently has the
geometry and electronic-structure of an isolated dislocation been
calculated;\cite{Woodward2001,Woodward2002,Woodward2004} this despite
the rapid advances in computer hardware and \abinit\ methods.
Previous \abinit\ calculations were limited by the long-range strain
field of a dislocation which is incommensurate with periodic boundary
conditions; hence, only dislocation
dipoles\cite{Arias1994,Frederiksen2003} or
quadrapoles\cite{Bigger1992,Arias2000} had been computed.  The advent
of ``flexible'' or ``Green function'' boundary conditions---first
conceived by Sinclair~\et,\cite{Sinclair1978} later redeveloped for
crack propagation\cite{Thomson1992} and for dislocations and
dislocation kinks\cite{Rao1998}---made possible the relaxation of the
core geometry of an isolated dislocation.  For a review of \abinit\
methods applied to dislocations, see [\onlinecite{Woodward2005}].
Flexible boundary conditions accurately treat the long-range strain
field away from the defect by using the harmonic ideal lattice
response in the form of the lattice Green function.  The lattice Green
function determines the relaxed position of an atom given the force on
it and its neighbors.  Flexible boundary conditions have been used to
model cracks,\cite{Thomson1992,Canel1995} dislocations and kinks in
bcc metals with classical potentials,\cite{Rao2001,Yang2001}
cross-slip processes in fcc metals,\cite{Rao1999} isolated screw
dislocations in bcc metals and ordered intermetallics with
\abinit,\cite{Woodward2001,Woodward2002,Woodward2004} and even
vacancies and free surfaces;\cite{Tewary2004} for a review of flexible
boundary condition approaches to nanomechanics of defects, see
[\onlinecite{Ortiz1999}].

Flexible boundary conditions are limited by the accuracy of the
lattice Green function.  Many closed-form results are known for the
lattice Green functions of cubic lattices with nearest neighbor
interactions.\cite{Tewary1973,MacGillivray1983} While the lattice
Green function is intimately related to the elastic constants and
dynamical matrix of a crystal, it has previously been computed for
realistic potentials from relaxation of atom positions given an
applied force.\cite{Sinclair1978} Rao~\et\ employed a ``direct
displacement'' technique where separate relaxation calculations in a
two-dimensional slab are used to numerically evaluate the lattice
Green function for short range, while switching to the known
long-range behavior of the elastic Green function.\cite{Rao1998}
Woodward~\et\ used this same technique with \abinit\ for Mo, Ta, and
TiAl, and found the lattice Green function matched the long-range
behavior at distances of only 5\AA, despite long-range metallic
bonding.\cite{Woodward2001} However, this technique is dependent on
the defect geometry---a lattice Green function computed for a
$[110]/2$ fcc screw dislocation cannot be used for the fcc edge
dislocation with a threading direction of $[1\bar 12]/2$.  Moreover,
relying on atomic relaxation can be prone to error in
density-functional methods when the applied forces become small.  A
more accurate and efficient approach instead relies on the dynamical
matrix and elastic constants, which can be computed using standard
techniques.

What follows is a general and accurate method for the computation of
the lattice Green function applicable for use in \abinit\ for a
variety of defect geometries.  In addition, we present and test an
estimate of the error in the lattice Green function due to the
geometry limitations of periodic-boundary conditions with \abinit.
Currently available methods for computing the dynamical matrix in
\abinit\ effectively produce a ``folded'' dynamical matrix, defined in
an artificial supercell---whether they rely on an finite supercell or
calculated on a discrete $k$-point
grid.\cite{Kunc82,Wei92,Frank95,Parlinski97,Baroni1987,Quong1992}
However, the interactions in \abinit\ have an unknown range, likely to
be larger than the artificial supercell.  What is required to compute
the lattice Green function is (1) a computational algorithm to
accurately use the limited dynamical matrix information, and (2) an
estimate of the error introduced from the dynamical matrix limitation.

\sect{harmonic} reviews the harmonic response functions in
a lattice---the dynamical matrix, and lattice Green function---and
relation to continuum elasticity theory.  \sect{lgf} derives the
general procedure for accurate numerical evaluation of the lattice
Green function, with specific application for zero-, one-, and
two-dimensional defects (point defects, dislocations, and boundaries,
respectively).  \sect{error} derives an error estimate for the lattice
Green function using only the dynamical matrix computation from a {\em
single} supercell and elastic constants.  The error estimate is
numerically tested using a simple-cubic lattice with random long-range
interactions, and is shown to be accurate even with supercells far
smaller than the interaction range.  Finally,
\sect{concl} concludes with discussion of applications to defect
calculations and the inherent length-scales in quasi-continuum methods
used in multiscale applications.

\section{Harmonic lattice response}
\label{sec:harmonic}

\bfig
\subfigure[][\ Local displacement produces forces (dynamical matrix)]{%
\includegraphics[width=\halfsmallfigwidth]{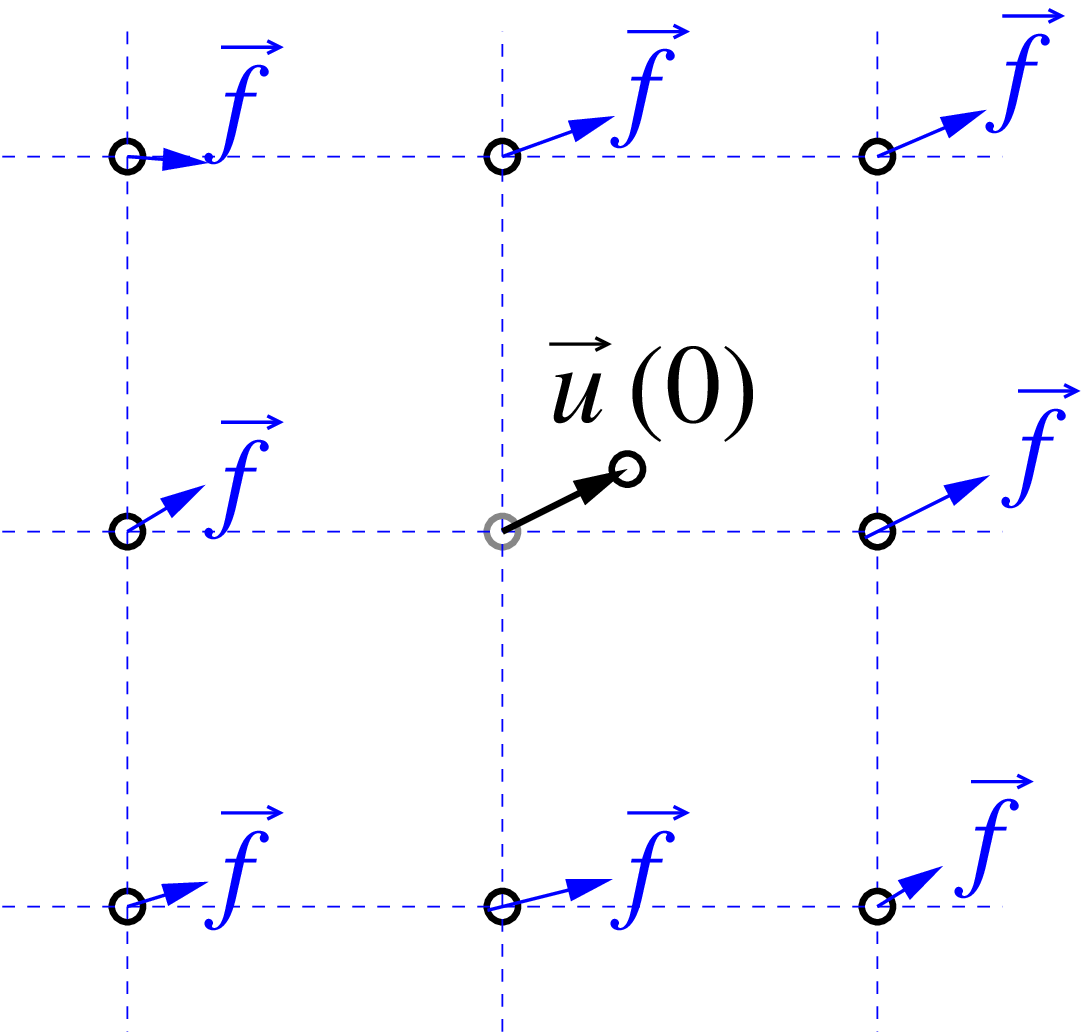}\ }
\subfigure[][\ Local force requires displacement to accommodate
             (lattice Green function)]{%
\ \includegraphics[width=\halfsmallfigwidth]{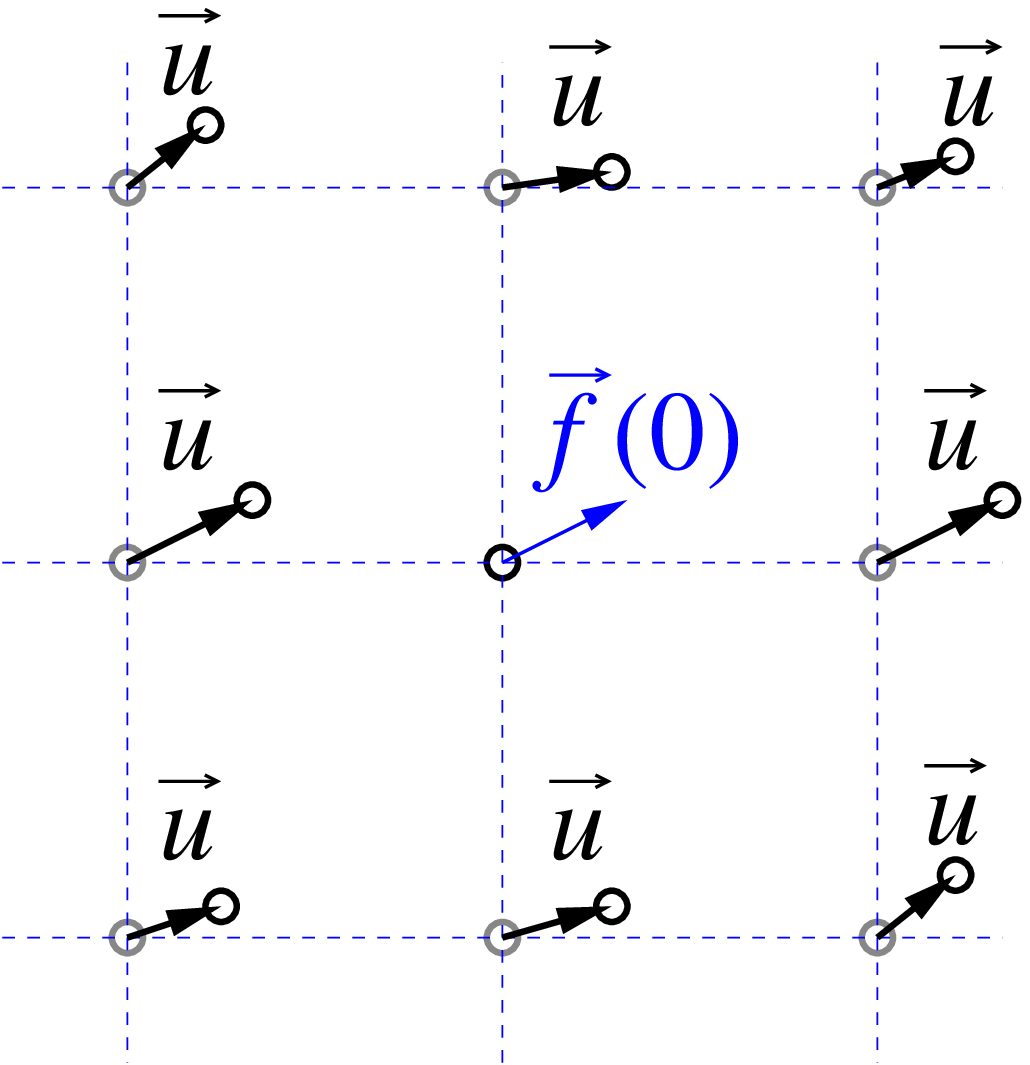}}

\caption{Schematic of harmonic lattice response.  Displacement of
  atoms $\uv$ in a crystal produce forces $\fv$ on neighboring atoms;
  the forces are given by the dynamical matrix $\D{}$ for small
  displacements.  Conversely, if an atom experiences a force,
  neighboring atoms must displace in order to accommodate the force;
  the displacement is given by the lattice Green function $\GL{}$ for
  small forces.}
\label{fig:lattice}
\efig

When atoms in a crystal are subject to applied or internal forces,
they respond by displacing from their ideal lattice sites; and
conversely, displacement from the ideal lattice sites produces
internal forces on atoms.  In the case of small displacements and
forces, the atoms in the lattice respond \term{harmonically}.
Harmonic response is characterized by a linear relationship between
displacement and force, given by two different lattice functions: the
\term{dynamical matrix}, and \term{lattice Green function}
(\fig{lattice} shows the responses schematically).  Below we review
their definitions and connections to anisotropic elasticity theory.
To simplify the notation, we assume a single atom Bravais lattice;
however, the approach translates readily to multiple atom Bravais
lattices.  Ionic crystals have additional complexities that are not
addressed here.%
\footnote{See [\protect\onlinecite{BornHuang1954}] for a review of phonons in
ionic crystals.}

The dynamical matrix is well known from the classical and quantum
theory of the harmonic crystal.\cite{BornHuang1954,Ashcroft76} Let
$\Rv$ and $\Rv'$ be two lattice sites in a crystal, and $\uv(\Rv)$ and
$\uv(\Rv')$ the displacements of the atoms from their ideal lattice
sites.  Then we write the harmonic potential energy as
\beu
U^\textrm{harm} = \frac{1}{2}\sum_{\Rv\Rv'} \uv(\Rv)\D{}(\Rv-\Rv')\uv(\Rv'),
\eeu
where $\D{}(\Rv-\Rv')$ is a $3\x 3$ matrix defined at lattice sites,
and the double sum ranges over all lattice sites.  For small
displacements, the harmonic potential energy will equal the total
potential energy $U^{\textrm{total}}$ of the system up to a constant
if the dynamical matrix components are
\beu
\D{ab}(\Rv-\Rv') = \left.\frac{\partial^2 U^\textrm{total}}{\partial
u_a(\Rv)\partial u_b(\Rv')}\right|_{\uv=\zero}.
\eeu
In an infinite bulk lattice, there are several symmetry relations.
First, $\D{}$ is a function only of $\Rv-\Rv'$, not $\Rv$ and $\Rv'$
independently due to translational symmetry.  Furthermore,
$\D{ab}(\Rv)=\D{ab}(-\Rv)=\D{ba}(\Rv)$ due to inversion symmetry and
independence of differentiation order.  Finally, if we displace every
atom identically, the total energy must remain a constant; hence, the
sum rule $\sum_{\Rv} \D{}(\Rv)=\zero$.  This sum rule has important
consequences for the lattice Green function.

The dynamical matrix linearly relates internal displacements and
forces, and connects elastic strain to elastic stress through
the elastic stiffness tensor.  Given displacements $\uv(\Rv)$ at atom
$\Rv$, the internal forces $\fv(\Rv')$ produced at atom $\Rv'$ are
given by
\be\label{eqn:Dij}
\fv(\Rv') = -\frac{\partial U^\textrm{harm}}{\partial \uv(\Rv')} = 
  -\sum_{\Rv} \D{}(\Rv-\Rv')\uv(\Rv).
\ee
An interesting special case of \eqn{Dij} are displacements
corresponding to a constant strain: $\uv(\Rv) = \CART{\eps}\Rv$, where
$\CART{\eps}$ is the strain tensor.  The crystal response is a
constant stress tensor $\CART{\sigma}$ which is linearly proportional
to the strain by Hooke's law: $\CART{\sigma} =
\C\CART{\eps}$.  This relationship is valid for small strains, and
defines the fourth-rank elastic stiffness tensor $\C$.  \eqn{Dij}
gives a connection between the elastic constants $C_{abcd}$ and the
dynamical matrix,
\be\label{eqn:longwaves}
-\sum_{\Rv} \D{ab}(\Rv)R_cR_d = V(C_{acbd}+C_{adbc}),
\ee
where $V$ is the volume of the unit cell.  \eqn{longwaves} can also be
derived using the method of long-waves.\cite{BornHuang1954} Hence, the
elastic constants contain information about long-range behavior of the
dynamical matrix.

The lattice Green function%
\footnote{This is the {\em static} bulk lattice Green function.  The
lattice Green function can be defined as a function of both time and
space to model phonon propagation.  In addition a defect lattice Green
function can be computed using the Dyson equation.  There, the
starting point is the bulk lattice Green function.}
linearly relates the forces on each atom to its displacement from the
ideal lattice.  That is,
\be\label{eqn:LGF}
\uv(\Rv') = -\sum_{\Rv} \GL{}(\Rv-\Rv')\fv(\Rv),
\ee
where $\GL{}(\Rv-\Rv')$ is the lattice Green function.  It obeys similar
symmetries to the dynamical matrix: $\GL{ab}(\Rv) = \GL{ab}(-\Rv) =
\GL{ba}(\Rv)$.  However, there is no sum rule for the lattice Green
function.%
\footnote{In fact, the sum of the lattice Green function over all
lattice sites diverges; this requires that the sum of all forces on
an infinite crystal body must vanish.}
At first, \eqn{LGF} may not appear useful; to compute the forces in
the {\em harmonic} potential, the displacements $\uv(\Rv)$ must
already be known.  However, if instead the forces on atoms in a
simulation are computed using the total energy $U^\mathrm{total}$
which is a function of relative atom positions, \eqn{LGF} allows one
to relax the atoms to their ideal lattice positions.  In that case,
the displacements are {\em not} known when computing the forces.  In
particular, the lattice Green function is used to create flexible
boundary conditions\cite{Sinclair1978,Rao1998} where an isolated
defect is surrounded by atoms that respond as if they are coupled to
infinite bulk.  This gives an accurate treatment of the long-range
stress field of a defect (such as a dislocation) while using forces
from $U^\mathrm{total}$ close to the defect.

As the long-range behavior of the dynamical matrix is connected to the
elastic constants, the long-range behavior of the lattice Green
function is connected to the elastic Green function.  The elastic
Green function $\GE{}$ is a continuum function that relates a
force-field $\fv(\xv)$ to the displacement-field $\uv(\vec y)$:
\beu
\uv(\vec y) = -\iiint d^3x\;\GE{}(\xv - \vec y)\fv(\xv).
\eeu
The elastic Green function can be computed knowing only the elastic
response of the continuum---i.e., the elastic constant tensor
$\C$.\cite{Stroh1962,Bacon1979} $\GE{}(\xv)$ satisfies the partial
differential equation
\be\label{eqn:EGF}
\sum_{abc} C_{iabc}\nabla_a\nabla_b \GE{cj}(\xv) =
  -\delta_{ij}\delta(\xv),
\ee
where $\delta(\xv)$ is the Dirac delta-function.  The lattice Green
function must match the elastic Green function as $\Rv\to\infty$, {\em
regardless} of how long-ranged the dynamical matrix is.

Lastly, the dynamical matrix and lattice Green function are inverses
of each other.  Substituting \eqn{Dij} into \eqn{LGF} gives
\be\label{eqn:inverse}
\sum_{\Rv'} \GL{}(\Rv - \Rv') \D{}(\Rv') = \ident\delta(\Rv),
\ee
where $\delta(\Rv)$ is the Kronecker delta-function.  \eqn{inverse} is
not strictly solvable because $\D{}$ is singular due to the sum rule.
The singularity is due to the lack of forces from a uniform
displacement of all atoms; thus, the displacements from \eqn{LGF} will
be known only up to a constant shift in the entire lattice.  This
overall translational symmetry in the lattice Green function provides
for the ``flexibility'' in flexible boundary conditions: bulk lattice
response can be simulated {\em without} specifying an origin for the
lattice.  Mathematically, the singularity in $\D{}$ must be carefully
treated to compute the lattice Green function accurately.

The computation of the lattice Green function is more tractable in
reciprocal space.  The lattice functions can be written as periodic
functions of vectors $\kv$ in the Brillouin zone (BZ) of
reciprocal space,\cite{Ashcroft76}
\beu
\GLk{}(\kv) = \sum_{\Rv} e^{i\kv\cdot\Rv} \GL{}(\Rv)
,\quad
\GL{}(\Rv) = V \intBZ \frac{d^3k}{(2\pi)^3}\; 
  e^{-i\kv\cdot\Rv}\GLk{}(\kv).
\eeu
In reciprocal space, the inverse equation \eqn{inverse} simplifies to
$\GLk{}(\kv)\Dk{}(\kv) = \ident$ for all $\kv$.  The singularity of
$\D{}$ is reduced to the gamma point $\kv=0$, where $\Dk{}(0) =
\zero$; for all other points, $\GLk{}(\kv) = \inv{[\Dk{}(\kv)]}$.  The
inverse is well-defined for metastable crystal structures; i.e.,
crystal structures without unstable phonon modes.

The computation of lattice Green function relies on accurate
computation of the dynamical matrix.  While computing the
dynamical-matrix is straightforward for interactions with a finite
cutoff, it is difficult for \abinit\ methods which may have long-range
interactions (such as Friedel-oscillations).  Two methods have
emerged: direct force\cite{Kunc82,Wei92,Frank95,Parlinski97} and
linear-response.\cite{Baroni1987,Quong1992} Both methods compute the
reciprocal-space dynamical matrix on a discrete grid of k-points in
the BZ.  This is equivalent to folding the real-space dynamical matrix
into an artificial supercell.  We use the folded dynamical matrix to
compute the lattice Green function; thus, we need to evaluate the
effect of the cutoff on the accuracy of the resulting lattice Green
function.  We do so with the elastic constants, which can be found
{\em separately} by computing the response of a periodic cell to
uniform strains.  \eqn{longwaves} relates the elastic constants to the
long-range behavior of the true dynamical matrix.  This relation
provides an estimate for the deviation of the long-range elastic Green
function from the lattice Green function, which is turn gives an error
estimate for using the folded dynamical matrix.  More importantly,
this estimate does {\em not} rely on a convergence test computation
comparing increasingly larger supercells.

Finally, it should be noted that the lattice Green function defined in
\eqn{LGF} can be modified for different bulk boundary conditions.
\eqn{LGF} defines $\GL{}$ in infinite bulk, called the 3D 
lattice Green function, and it is useful for computation of point
defects.  If the forces and displacements have periodicity along a
lattice vector $\tv$, such as in a single straight dislocation defect,
the 2D lattice Green function is used: $\sum_n \GL{}(\Rv +n\tv)$.
Finally, if forces and displacements have periodicity along two
lattice vectors $\tv_1$ and $\tv_2$, such as in surfaces, grain
boundaries and interfaces, the 1D lattice Green function is used:
$\sum_{mn}
\GL{}(\Rv + m\tv_1 +n\tv_2)$.  Despite the simple summations used to
define the 2D and 1D lattice Green functions from the 3D, the sums
converge conditionally.  It should be remembered that the
``dimensionality'' of the lattice Green functions refer to the degrees
of freedom for the lattice vector $\Rv$---$\GL{}$ remains a $3\x 3$
matrix in all cases.  The dimensionality of the defect (0, 1, or 2)
plus the dimensionality of the lattice Green function (3, 2, or 1)
sums to 3.

\section{Computation of Lattice Green function}
\label{sec:lgf}

The procedure for numerical computation of the lattice Green function
separates the Fourier transform into pieces which can be inverse
Fourier transformed accurately.  The straightforward approach would be
to discrete inverse Fourier transform the inverse of the dynamical
matrix; however, this transform converges very slowly with increased
grid spacing due to the second-order pole at the gamma point.  The
inversion of the dynamical matrix to compute the lattice Green
function is still best performed in reciprocal space, where the large
$R$ behavior is exactly contained in the pole at $k=0$.  To accurately
compute the lattice Green function requires an analytic treatment of
the small $k$ behavior separated from the rest of the Brillouin zone.

The separation of the lattice Green function allows the inverse
Fourier transform to converge by analytically treating the
second-order pole.  Moreover, the separation can be evaluated for any
dynamical matrix, and for any dimension.  The second-order pole in
$\GLk{}$ comes from the expansion of $\Dk{}(\kv)$ for small $k$,
\be
\begin{split}
\Dk{}(\kv) &= \sum_{\Rv} \D{}(\Rv) \exp(i\kv\cdot\Rv)\\
  &\approx \sum_{\Rv} \D{}(\Rv) \left[1 -
   \frac{1}{2}(\kv\cdot\Rv)^2 + \frac{1}{24}(\kv\cdot\Rv)^4\right]\\
  &= \sum_{cd}k_ck_d\left\{
     -\frac{1}{2}\sum_{\Rv} \D{}(\Rv)R_cR_d\right\}\\
  &\quad  -\sum_{cdef}k_ck_dk_ek_f\left\{
     -\frac{1}{24}\sum_{\Rv}\D{}(\Rv)R_cR_dR_eR_f\right\}.
\end{split}
\ee
The final expression is rewritten in terms of two functions of $\kv$
of different order in $k$: $k^2\Le(\hat k) - k^4\Lq(\hat k)$, where
$\hat k=\kv/k$.  We relate the first function $\Le(\hat k)$ to the
elastic constants by \eqn{longwaves},
\beu
\sum_{cd}k_ck_d\left\{-\frac{1}{2}\sum_{\Rv} \D{ab}(\Rv)R_cR_d\right\} = 
  V\sum_{cd} k_c C_{cabd} k_d,
\eeu
which gives $\Le(\hat k) = V[\hat k\C\hat k]$, where $\C$ is the
fourth-rank elastic stiffness tensor.  On the other hand, the quartic
function $\Lq(\hat k)$ has no similar simple connection.  With the
definitions of $\Le$ and $\Lq$, the lattice Green function expands for
small $k$ as
\beu
\begin{split}
\GLk{}(\kv) &= \inv{[\Dk{}(\kv)]}\\
  &= \inv{[k^2\Le(\hat k) - k^4\Lq(\hat k) + O(k^6)]}\\
  &= k^{-2}\inv{[\Le(\hat k)]}\inv{[\ident - k^2\Lq(\hat k)\inv{[\Le(\hat
       k)]} + O(k^4)]}\\
  &= k^{-2}\inv{[\Le(\hat k)]} + \inv{[\Le(\hat k)]}\Lq(\hat k)\inv{[\Le(\hat
       k)]} + O(k^2)\\
  &\equiv \GEk{}(\kv) + \Gdck{}(\kv) + O(k^2),
\end{split}
\eeu
where
\be
\label{eqn:GEk}
\begin{split}
\GEk{}(\kv) &\equiv k^{-2}\inv{[\Le(\hat k)]}\\
&=\frac{1}{Vk^2}\inv{[\hat k\C\hat k]},
\end{split}
\ee
and
\be
\label{eqn:Gdck}
\begin{split}
\Gdck{}(\kv) &\equiv \inv{[\Le(\hat k)]}\Lq(\hat k)\inv{[\Le(\hat k)]} \\
&=
   \GEk{}(\hat k)
   \left[-\frac{1}{24}\sum_{\Rv}\D{}(\Rv)(\hat k\cdot\Rv)^4\right]
   \GEk{}(\hat k).
\end{split}
\ee
The first function $\GEk{}$ is the second-order pole at the gamma
point, which is the Fourier transform of the elastic Green function.
The second function $\Gdck{}$ is independent of $|k|$, representing a
discontinuity at the gamma point in the lattice Green function that
appears only after the second-order pole is subtracted out; this
function is called the discontinuity correction.

This expansion is used to separate the Fourier transform of the
lattice Green function in the entire Brillouin zone into three pieces:
the elastic Green function, discontinuity correction, and
semicontinuum correction.  We introduce the continuous and
differentiable cutoff function $\fc(k/\kmax)$ with parameter
$0<\alpha<1$,
\be
\label{eqn:cutoff}
\fc(x) =
\begin{cases}
  1 & : 0\le x<\alpha\\
  3(\frac{1-x}{1-\alpha})^2 - 2(\frac{1-x}{1-\alpha})^3 & : \alpha\le x<1\\
  0 & : 1\le x\\
\end{cases}
,
\ee
where $\kmax$ is the radius of a sphere inscribed in the Brillouin
zone.  While final evaluation of $\GL{}(\Rv)$ is independent of
$\alpha$, all computations to follow use $\alpha=1/2$.  Then, the
semicontinuum correction is defined for $\kv$ in the first Brillouin
zone as
\be\label{eqn:semicont}
\Gsck{}(\kv) = \inv{[\Dk{}(\kv)]} - (\GEk{}(\kv) + \Gdck{}(\kv))\fc(k/\kmax).
\ee
The elimination of the second-order pole at the gamma-point by using a
cutoff version of the elastic Green function is related to the
semicontinuum method of Tewary.\cite{Tewary1971} However, his
semicontinuum approach used a Gaussian cutoff which does not vanish at
the Brillouin zone edge, and does not treat the discontinuity produced
at the gamma point.  The final lattice Green function is the sum of
three pieces: elastic Green function, discontinuity correction, and
semicontinuum correction.

\bfigwide
\centering
\includegraphics[width=\wholefigwidth]{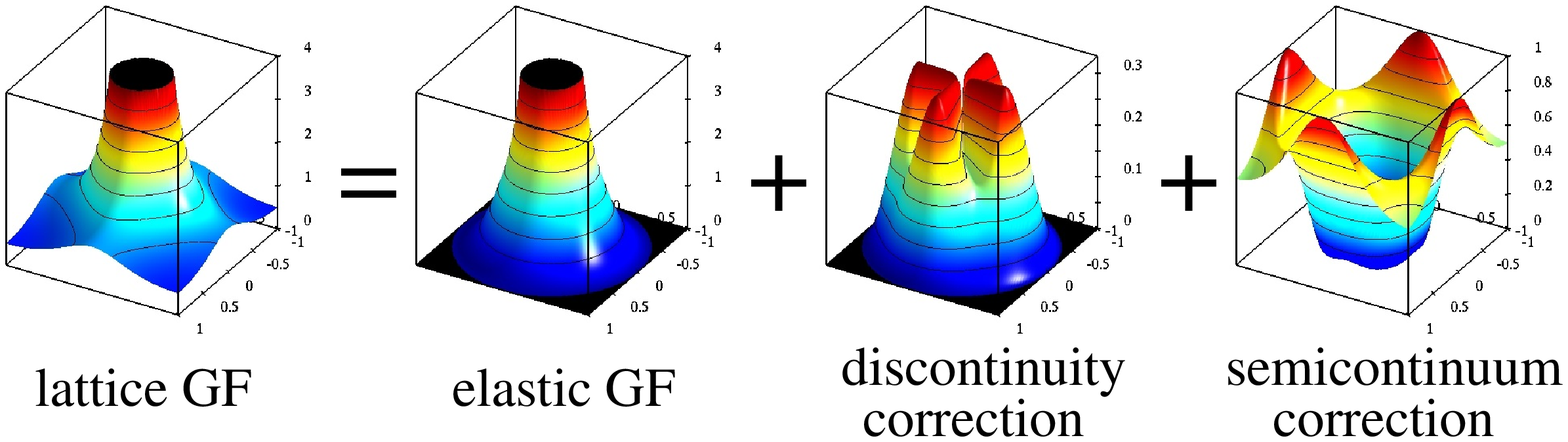}

\caption{Separation of lattice Green function for a square lattice in
  2-dimensional reciprocal space into elastic Green function,
  discontinuity correction, and semicontinuum correction (note
  different vertical scales).  The lattice Green function has the
  periodicity of the reciprocal lattice, and a second-order pole at
  the gamma point.  The elastic Green function scales as $k^{-2}$, and
  is cutoff to smoothly go to zero at the Brillouin zone edges.  The
  removal of the second-order pole creates a discontinuity independent
  of $|k|$ at the gamma point; the discontinuity correction removes
  the discontinuity and smoothly goes to zero at the Brillouin zone
  edges.  The remaining difference between the lattice Green function
  and the first two terms is the semicontinuum correction, which is
  smooth everywhere in the Brillouin zone.}
\label{fig:lgf}
\efigwide

\fig{lgf} shows an example of the separation of the lattice
Green function into the three terms for a square lattice.  The lattice
Green function shown comes from a square lattice with lattice constant
$a_0=\pi$ and nearest-neighbor interactions.  For this case,
$\GLk{}(k_x,k_y)= (\sin^2(\pi k_x/2) + \sin^2(\pi k_y/2))^{-1}$.  The
second-order pole at origin is given by the elastic Green function
$\GEk{}(k_x,k_y) = 4/(\pi|k|)^2$; it is multiplied by the cutoff
function with $\kmax = 1$ so as to vanish at the Brillouin zone edge.
Subtracting the pole from the lattice Green function produces a
function with a discontinuity at the gamma point.  The discontinuity
at the origin is given by the discontinuity correction
$\Gdck{}(k_x,k_y) = (k_x^4 + k_y^4)/(3|k|^4)$ which is multiplied by
the cutoff function.  Subtracting the discontinuity produces the
semicontinuum correction, $\Gsck{}(k_x,k_y)$, given by \eqn{semicont}.

The evaluation of the lattice Green function in real space is
accomplished by inverse Fourier transforming the semicontinuum
correction $\Gsck{}$, the cutoff elastic Green function $\GEk{}\fc$,
and the cutoff discontinuity correction $\Gdck{}\fc$.  The
semicontinuum correction $\Gsck{}(\kv)$ is evaluated on a discrete
k-point grid in the Brillouin zone using \eqn{semicont}; inversion of
the dynamical matrix for small $k$ must be handled carefully to avoid
numerical noise.  A discrete inverse Fourier transform converges well
with grid spacing because $\Gsck{}$ is smooth throughout the Brillouin
zone.  The cutoff elastic Green function $\GEk{}\fc$ and discontinuity
correction $\Gdck{}\fc$ are expanded as functions of $\hat k$ using
spherical harmonics or a Fourier series depending on the
dimensionality of the problem.  In this form, their inverse Fourier
transforms can be analytically reduced to a one-dimensional integral
of non-singular functions over a finite range that is computed
numerically to the desired accuracy.  The details of this reduction
depends on the periodicity of the lattice Green function.
\tab{summary} gives a brief overview of the results, and
\tab{eqn-summary} gives a summary of the equations.

\begingroup
\squeezetable 
\begin{table}
\caption{Overview of lattice Green function computation for different
  dimensionality.  The dimensionality of the lattice Green function is
  determined by the type of defect being simulated: the defect
  dimensionality plus the lattice Green function dimensionality is
  three.  While the lattice Green function has the same form in
  reciprocal space, the periodicity determines the range of Brillouin
  zone integration, and the functions used to expand the $\hat k$
  dependence of the elastic Green function and discontinuity
  correction.  The range of BZ integration produces different large
  $R$ behavior for both the elastic Green function---also given by
  elasticity theory---and the discontinuity correction.  The
  one-dimensional case has no $\hat k$ dependence, so there is no
  angular expansion nor is a discontinuity correction required.}
\label{tab:summary}
  \begin{ruledtabular}
  \begin{tabular}{lccc}
    & \entry{3D} & \entry{2D} & \entry{1D}
    \\
  \tabpara{0.80in}{Defect type:\\(dimensionality)}
     & \tabparac{0.75in}{point (0D)}
     & \tabparac{0.75in}{dislocation, crack tip (1D)}
     & \tabparac{0.75in}{free surface, boundary (2D)}
     \\[6pt]\hline
  \tabpara{0.80in}{Brillouin zone integration:}
     & \tabparac{0.75in}{full BZ}
     & \tabparac{0.75in}{plane(s) $\perp$ to threading direction}
     & \tabparac{0.75in}{line(s) $\perp$ to surface plane}
     \\[6pt]\hline
  \tabpara{0.80in}{Angular expansion in Brillouin zone:}
     & \tabparac{0.75in}{spherical harmonics $Y_{lm}(\theta_k,\phi_k)$}
     & \tabparac{0.75in}{Fourier series in plane $e^{in\phi_k}$}
     & \tabparac{0.75in}{N/A}
     \\[6pt]\hline
  \tabpara{0.80in}{Large $R$ elastic Green function:} 
     & $R^{-1}$ & $-\ln R + R^0$ & $R$
     \\[6pt]\hline
  \tabpara{0.80in}{Large $R$ discontinuity correction:} 
     & $R^{-3}$ & $R^{-2}$ & N/A
     \\
  \end{tabular}
  \end{ruledtabular}
\end{table}
\endgroup

\subsection{3D lattice Green function: 0D defects}

To facilitate inverse Fourier transformation, the elastic Green
function $\GEk{}$ (\eqn{GEk}) and discontinuity correction $\Gdck{}$
(\eqn{Gdck}) are expanded as a spherical harmonic series, whose
coefficients are computed numerically.  The expansions
\beu
\GEk{}(\kv) = \frac{1}{k^2} \sum_{l=0}^{\Lmax}\sum_{m=-l}^l
              \GEk{lm} Y_{lm}(\hat k),\quad
\Gdck{}(\kv) = \sum_{l=0}^{\Lmax}\sum_{m=-l}^l
               \Gdck{lm} Y_{lm}(\hat k),
\eeu
are truncated for $l>\Lmax$; $\Lmax$ is chosen for each expansion so
that the $lm$ components above $\Lmax$ are less than $10^{-11}$ of the
largest $lm$ component below $\Lmax$.  Moreover, as both $\GEk{}$ and
$\Gdck{}$ are symmetric with respect to inversion, only even $l$
values are nonzero.  The normalized spherical harmonics are given by
\beu
Y_{lm}(\theta,\phi)
  = e^{im\phi} \sqrt{\frac{2l+1}{4\pi}\frac{(l-m)!}{(l+m)!}}
    P_l^m(\cos\theta),
\eeu
for $\hat k = (\sin\theta\cos\phi,\sin\theta\sin\phi,\cos\theta)$,
where $P_l^m(x)$ is the associated Legendre polynomial without the
$(-1)^m$ phase.\cite{MathFunctions} To compute the spherical harmonic
expansion, the elastic Green function and discontinuity correction are
evaluated on a spherical $N\x N$ grid $|\kv|=1$ given by $\phi_i=2\pi
i/N$, and $\theta_j = \arccos(u_j)$, where $u_j$ are the $N$ roots of
the $N^\mathrm{th}$ order Legendre polynomial $P_N(u)$.  This grid
allows the computation of expansion elements up to $\Lmax = N/2 - 1$.
The spherical harmonic components are evaluated by (1) discrete
Fourier transforming the $\phi_i$ grid to $m$ components, then (2)
using Gaussian-Legendre quadrature with weights $w_j$ on the
$\theta_j$ grid to produce $l$ components:
\beu
\GEk{lm} = \frac{2\pi}{N}\sum_{j=0}^{N-1} w_j 
  \sqrt{\frac{2l+1}{4\pi}\frac{(l-m)!}{(l+m)!}} P_l^m(u_j)
  \sum_{i=0}^{N-1} e^{-im\phi_i} \GEk{}(\theta_j, \phi_i),
\eeu
and identically for $\Gdck{}$.  As a final step, it is useful to
reduce numerical error in the expansion by explicitly enforcing the
point group symmetry of the lattice on the expansion; this is done
using the Wigner D-matrix,\cite{Altmann1963} modified to take into
account the effect of a symmetry operation on the $3\x 3$ matrix
elements.

Given the spherical harmonic series, inverse Fourier transformation
reduces to a single integral over a finite range.  Writing the inverse
Fourier transform integral in spherical harmonics over the BZ gives
\beu
\begin{split}
\GE{}(\Rv) &= \sum_{lm}^{\Lmax}\GEk{lm} \frac{V}{(2\pi)^3}
  \intkmax dk\; \fc(k/\kmax)\!
  \intsphere d^2\hat k\; 
     e^{ikR(\hat k\cdot\hat R)} Y_{lm}(\hat k),\\
\Gdc{}(\Rv) &= \sum_{lm}^{\Lmax}\Gdc{lm} \frac{V}{(2\pi)^3}
  \intkmax dk\; k^2\fc(k/\kmax)\!
  \intsphere d^2\hat k\; 
     e^{ikR(\hat k\cdot\hat R)} Y_{lm}(\hat k),\\
\end{split}
\eeu
by virtue of the cutoff function.  The double integral over $\hat k$
is evaluated analytically as \eqn{3dangularfinal}, so 
\beu
\begin{split}
\GE{}(\Rv) &= \sum_{lm}^{\Lmax} \GEk{lm}Y_{lm}(\hat R) (-1)^{l/2} \frac{V}{2\pi^2}
  \intkmax dk\; \fc(k/\kmax) j_l(kR),\\
\Gdc{}(\Rv) &= \sum_{lm}^{\Lmax} \Gdck{lm}Y_{lm}(\hat R) (-1)^{l/2} \frac{V}{2\pi^2}
  \intkmax dk\; k^2\fc(k/\kmax) j_l(kR),\\
\end{split}
\eeu
where $j_l(x)$ is the spherical Bessel function.

The finite one-dimensional integrals are smooth functions that can be
evaluated numerically to required accuracy.  For the special case of
$R=0$, 
\beu
\begin{split}
  \intkmax dk\; \fc(k/\kmax) j_l(0) 
    &= \delta_{l,0} \kmax \left[1 - \frac{1-\alpha}{2}\right],\\
  \intkmax dk\; k^2 \fc(k/\kmax) j_l(0) 
    &= \delta_{l,0} \kmax^3 \left[\frac{1}{3} -
	\frac{(1-\alpha)(2\alpha^2+5\alpha+8)}{30}\right].
\end{split}
\eeu
For $R\ne 0$, we define $f^{(0)}_l(x)$ and $f^{(2)}_l(x)$ as
\be
\label{eqn:3dintegrals}
\begin{split}
f^{(0)}_l(x) &\equiv \frac{2}{\pi}\intx du\; j_l(u)\fc(u/x),\\
f^{(2)}_l(x) &\equiv \frac{2}{\pi}\intx du\; u^2j_l(u)\fc(u/x),
\end{split}
\ee
so
\beu
\begin{split}
  \frac{V}{2\pi^2}\intkmax dk\; \fc(k/\kmax) j_l(kR) 
    &= \frac{V}{4\pi R}f^{(0)}_l(\kmax R),\\
  \frac{V}{2\pi^2}\intkmax dk\; k^2 \fc(k/\kmax) j_l(kR) 
    &= \frac{V}{4\pi R^3}f^{(2)}_l(\kmax R).
\end{split}
\eeu
The $f_l$ functions are evaluated numerically by splitting the
integrals into intervals between roots of $j_l(x)$, and then using the
QAG adaptive integration algorithm with 61 point Gauss-Kronrod rules
from \textsc{quadpack}.\cite{QUADPACK} An important limiting case is
for $R\to\infty$ where the functions can be evaluated analytically.
From \INT\ expression 6.561.14 gives for even $l$
\beu
\begin{split}
\lim_{x\to\infty} f^{(0)}_l(x) &= 
  \prod^{l}_{k\textrm{\ odd}} k \left/ 
  \prod^{l}_{k\textrm{\ even}} k\right.,\\
\lim_{x\to\infty} f^{(2)}_l(x) &= (l+1)l
  \prod^{l}_{k\textrm{\ odd}} k \left/ 
  \prod^{l}_{k\textrm{\ even}} k\right.
.
\end{split}
\eeu
These results are summarized in \tab{eqn-summary}.

The inverse Fourier transform of the semicontinuum correction
$\Gsck{}$ is performed with a discrete transform on a grid in the
Brillouin zone.  There are different techniques for constructing a
k-point mesh,\cite{Chadi73,Monkhorst76} but a uniform grid of
$\kv$-points centered at the gamma point inside the BZ suffices.  The
primary requirement is that each k-point lie in the first BZ; and in
that way, the points given by $|k|<\kmax$ form a sphere.  The spacing
of the grid is determined by the largest magnitude lattice vector
$R_\textrm{max}$ in the desired domain of $\GL{}(\Rv)$.  To avoid
aliasing errors, the grid spacing $\Delta k$ must be smaller than
$2\pi / R_\textrm{max}$, though a smaller spacing is preferable.  For
large $R$, substituting the elastic Green function for the lattice
Green function introduces only small errors, hence reducing the
effective $R_\textrm{max}$ and k-point mesh that is used.  We estimate
the deviation in detail in \sect{error}.

\subsection{2D lattice Green function: 1D defects}

The introduction of a threading direction reduces the lattice Green
function to a two-dimensional slab and modifies the inverse Fourier
transformations.  The forces and displacements of atoms around a
dislocation line or a crack tip have a periodicity given by a
threading lattice vector $\tv$.  The periodicity is represented in the
lattice Green function by the 2D lattice Green function, $\sum_n
\GL{}(\Rv+n\tv)$.  As with the 3D lattice Green function, evaluation
of the 2D lattice Green function is best performed in Fourier space,
and inverse Fourier transforming to real space.  Then,
\be\label{eqn:2dLGF}
\begin{split}
\CART{G}^{\textrm{L-2D}}(\Rv)
  & = \sum_{n=-\infty}^{\infty} \CART{G}^{\textrm{L-3D}}(\Rv + n\tv) \\
  &= \sum_{n=-\infty}^{\infty} \frac{V}{(2\pi)^3} \intBZ d^3k\;
     e^{-i\kv\cdot\Rv} e^{-in\kv\cdot\tv } \GLk{}(\kv) \\
  &= \sum_{\kpara \in \textrm{BZ}}
     \frac{V}{|\tv\,|}\intBZtwo\frac{d^2k_\perp}{(2\pi)^2}\;
     e^{-i(\kperp+\kpara)\cdot\Rv} \GLk{}(\kperp+\kpara)
\end{split}
\ee
where the (finite) summation is over $\kpara = 2\pi m\tv /|\tv\,|^2$
($m$ integer) that are inside the BZ, and two-dimensional integration
is over $\kperp$ that are perpendicular to $\tv$ and inside the BZ.
This is by virtue of the summation over $n$ which produces a delta
function on $\exp(i\kv\cdot\tv)-1$.  \eqn{2dLGF} still has a pole in
$\GLk{}$ to contend with, but it lies purely in the plane of $\kpara =
\zero$.  Hence, for $\kpara\ne \zero$, the value of
$\GLk{}=\inv{[\Dk{}]}$ is used, and a discrete inverse Fourier
transform is performed.  Then, the remaining difficulty is the
$1/k_\perp^2$ pole at the gamma point in the 2D inverse Fourier
transform.

The pole at the gamma point in 2D is split into three contributions
for inverse Fourier transformation: elastic Green function,
discontinuity correction, semicontinuum correction.  The elastic Green
function $\GEk{}$ and discontinuity correction $\Gdck{}$ are expanded
as a truncated Fourier series in the plane of $\kperp$,
\beu
\GEk{}(\kperp) = \frac{1}{k_\perp^2} \sum_{n=0}^{\Nmax} \GEk{n} e^{in\phi_k},\quad
\Gdck{}(\kperp) = \sum_{n=0}^{\Nmax} \Gdck{n} e^{in\phi_k},
\eeu
where $\phi_k$ is the angle of $\kperp$ relative to an (arbitrary)
normalized in-plane reference direction $\nhat$ ($\nhat\cdot\tv = 0$).
The truncation $\Nmax$ is chosen for each expansion so that the $n$
components above $\Nmax$ are less than $10^{-11}$ of the largest $n$
component below $\Nmax$.  Since both $\GEk{}$ and $\Gdck{}$ have
inversion symmetry, only even $n$ values are nonzero.  The Fourier
series components are evaluated by computing $\GEk{}(\kv)$ on a $N$
element circular grid $|\hat k_\perp|=1$ at a series of angles
$\phi_i = 2\pi i/N$ relative to $\nhat$.  The discrete Fourier
transform gives
\beu
\GEk{n} = \frac{1}{N}\sum_{i=0}^{N-1} e^{-in\phi_i}
  \GEk{}(\hat k_\perp(\phi_i)),
\eeu
and identically for $\Gdck{}$.  Note that $\GEk{}(\kv)$ and
$\Gdck{}(\kv)$ are the same functions that appear in the 3D lattice
Green function (given by \eqn{GEk} and \eqn{Gdck}); for the 2D lattice
Green function, they are only evaluated in the plane through the gamma
point.

Given the Fourier series, inverse Fourier transformation reduces to a
single integral over a finite range.  The $\kpara\ne\zero$ terms of
\eqn{2dLGF} have no singularities, so they can be evaluated
numerically using a discrete inverse Fourier transform with a discrete
grid for $\kperp$, where construction of this grid is described below.
Hence, the elastic Green function and discontinuity corrections are
only evaluated for $\kpara=\zero$.  The inverse Fourier transform
integral over the BZ gives
\beu
\begin{split}
&\GE{}(\Rv) \;[\kpara=\zero] = \\
&\sum_{n=0}^{\Nmax} \GEk{n} \frac{V}{|\tv\,|(2\pi)^2}
  \intkmax \frac{dk}{k}\;\fc(k/\kmax)\intcirc d\phi_k\;
  e^{ik \Rperp \cos(\phi_k-\phi_R)} e^{in\phi_k},
\end{split}
\eeu
where $\Rperp = \sqrt{R^2 - (\Rv\cdot\tv)/t^2}$ is the magnitude of
$\Rv$ perpendicular to $\tv$, and $\phi_R$ is the in-plane angle of
$\Rv$ relative to $\nhat$ ($\phi_R=\arccos((\nhat\cdot\Rv)/\Rperp)$).
The integral over $\phi_k$ is given by expression 8.411.1 in \INT\ as
$2\pi(-1)^{n/2} J_n(k\Rperp)\exp(in\phi_R)$ where $J_n(x)$ is the
Bessel function, so
\beu
\begin{split}
\GE{}(\Rv) &= \sum_{n=0}^{\Nmax} \GEk{n}e^{in\phi_R} (-1)^{n/2}
  \frac{V}{|\tv\,|2\pi}
  \intkmax dk\; k^{-1}\fc(k/\kmax) J_n(k\Rperp),\\
\Gdc{}(\Rv) &= \sum_{n=0}^{\Nmax} \Gdck{n}e^{in\phi_R} (-1)^{n/2}
  \frac{V}{|\tv\,|2\pi}
  \intkmax dk\; k \fc(k/\kmax) J_n(k\Rperp).
\end{split}
\eeu

The finite one-dimensional integrals are smooth functions for $n>0$
that can be evaluated numerically to required accuracy.  For the
special case of $\Rperp=0$ and $n\ne 0$, the integrals over $k$ are
zero.  For $n\ne 0$ and $\Rperp\ne 0$, we define $F^{(0)}_n(x)$ and
$F^{(2)}_n(x)$ as
\be
\label{eqn:2dintegrals}
\begin{split}
F^{(0)}_n(x) &\equiv \intx du\; u^{-1}J_n(u)\fc(u/x),\\
F^{(2)}_n(x) &\equiv \intx du\; uJ_n(u)\fc(u/x).
\end{split}
\ee
so
\beu
\begin{split}
  \frac{V}{2\pi}\intkmax dk\; k^{-1}\fc(k/\kmax) J_n(k\Rperp)
    &= \frac{V}{2\pi}F^{(0)}_n(\kmax \Rperp),\\
  \frac{V}{2\pi}\intkmax dk\; k\fc(k/\kmax) J_n(k\Rperp)
    &= \frac{V}{2\pi \Rperp^2}F^{(2)}_n(\kmax \Rperp).\\
\end{split}
\eeu
As in the 3D case, the $F_n$ functions for $n\ne0$ are evaluated
numerically by splitting the integrals into intervals between roots of
$J_n(x)$, and then using the QAG adaptive integration algorithm with
61 point Gauss-Kronrod rules from \textsc{quadpack}.\cite{QUADPACK}
For $u<10^{-5}$ in the integral, using $J_n(u)/u\approx 1/2 (u/2)^{n-1}$
eliminates the division by zero.  An important limiting case is for
$\Rperp\to\infty$ where the functions can be evaluated analytically.
Expressions 6.561.14 and 6.621.4 in \INT\ give for $n\ne 0$
\beu
\lim_{x\to\infty} F^{(0)}_n(x) = \frac{1}{n},\quad
\lim_{x\to\infty} F^{(2)}_n(x) = n.
\eeu

\bfig
\centering
\includegraphics[width=2.5in]{F0.eval.eps}

\caption{Evaluation of the $R$ scaling for the circularly symmetric
  portion of the 2D elastic Green function given by $F^{(0)}_0(\kmax
  R)$, with $\alpha=1/2$ and $\kmax=\pi$.  The long range behavior of
  the elastic Green function in two dimensions scales as $-\ln(R)$.
  The cutoff function retains the correct large $R$ behavior, with
  small deviations after $R=2$ in lattice units.  Moreover, the cutoff
  removes the divergence at the origin.}
\label{fig:F0eval}
\efig

The finite one-dimensional integrals for the $n=0$ case require
additional analytic manipulation to be evaluated numerically.
\fig{F0eval} shows the convergence to the long-range behavior for
$F^{(0)}_0(\kmax \Rperp)$.  The function $F^{(2)}_0(x)$ is
well-behaved for all values of $x$, and the limiting case of
$x\to\infty$ is 0 as given above.  The $\Rperp=0$, $n=0$ integral for
the discontinuity correction is
\beu
  \intkmax dk\; k\fc(k/\kmax) J_0(0) = \kmax^2 \left[
  \frac{1}{2} - \frac{(1-\alpha)(3\alpha+7)}{20} \right].
\eeu
The $n=0$ integral for the elastic Green function does not converge as
written, because the limit as $\Rperp\to\infty$ diverges.  To perform
the integration, it is useful to remember from elasticity theory that
the 2D elastic Green function in real space scales as $\ln(\Rperp)$.
The Fourier transform of $\ln|\vec r|$ in two-dimensions is
\beu
\frac{1}{2\pi}\iint\negthickspace d^2r\;e^{i\kv\cdot\vec r} \ln|\vec r|
  = \Int{0}{\infty}dr\;r\ln(r) J_0(kr)
  = -\frac{1}{k^2},
\eeu
which means the inverse Fourier transform integral is
\beu
\frac{1}{2\pi}\iint\negthickspace d^2k\;e^{-i\kv\cdot\vec r}\frac{1}{k^2}
  = \Int{0}{\infty}\frac{dk}{k}\;J_0(kr) = -\ln|r|.
\eeu
Using this relation, the pole at $k=0$ can be evaluated analytically as
\beu
\begin{split}
& \intkmax \frac{dk}{k}\;\fc(k/\kmax)J_0(k\Rperp)\\
&=  \Int{0}{\infty} \frac{dk}{k}\;J_0(k\Rperp)
    +\Int{\alpha\kmax}{\infty}\frac{dk}{k}\;J_0(k\Rperp)
    \left[\fc(k/\kmax)-1\right]\\
&= -\ln(\Rperp) 
    -\Int{\alpha\kmax}{\infty}\frac{dk}{k}\;J_0(k\Rperp)
    +\Int{\alpha\kmax}{\kmax}\frac{dk}{k}\;J_0(k\Rperp)\fc(k/\kmax)\\
&= \ln\left(\frac{\alpha\kmax}{2}\right) + \gamma
 -\frac{1}{2}\sum_{n=0}^{\infty}\frac{(-1)^n}{(n+1)(n+1)!^2}
     \left(\frac{\alpha\kmax \Rperp}{2}\right)^{2n+2}\\
&\quad + \Int{\alpha\kmax}{\kmax}\frac{dk}{k}\;J_0(k\Rperp)\fc(k/\kmax),
\end{split}
\eeu
where $\gamma$ is the Euler constant ($\gamma\approx 0.5772156649$).
The remaining integral is evaluated numerically as before, and the
series is summed numerically to within $10^{-11}$.  The expression is
finite for $\Rperp=0$, and in the limit $\Rperp\to\infty$ becomes
$-\ln(\Rperp)$ (c.f.~\fig{F0eval}).  The $\Rperp=0$, $n=0$ integral
for the elastic Green function is
\beu
\begin{split}
  &\intkmax \frac{dk}{k}\; \fc(k/\kmax) J_0(0) = \ln(\kmax) + [\gamma-\ln2] \\
  &\quad + \frac{6\alpha^2(3-\alpha)\ln\alpha
     -(1-\alpha)(5\alpha^2 -22\alpha + 5)}{6(1-\alpha)^3}.
\end{split}
\eeu
These results are summarized in \tab{eqn-summary}.

The inverse Fourier transform of the semicontinuum correction
$\Gsck{}$ is performed with a discrete transform on a grid lying on
planes in the Brillouin zone.  The planes are specified by the
threading direction in the lattice $\tv$; to form a planar grid
requires two in-plane lattice vectors $\nperp$ and $\mperp$.  All
three vectors are mutually perpendicular, though not normalized.  The
$N\x M$ grid is the combination of $\kpara$ and $\kperp$, with
\beu
\kv(t,n,m) = \frac{2\pi\tv}{|\tv\,|^2}t + 
  \frac{2\pi\nperp}{|\nperp|^2}\frac{n}{N} + 
  \frac{2\pi\mperp}{|\mperp|^2}\frac{m}{M},
\eeu
where $t,n,m$ are integers ranging over the interior of the BZ.  The
integers $N$ and $M$ specify the in-plane grid spacing, and must be
chosen sufficiently large to remove aliasing effects out to
$R_\textrm{max}$.  As for the 3D lattice Green function, the deviation
between the 2D lattice Green function and 2D elastic Green function
decreases with distance, thus requiring the computation of the lattice
Green function out to a fixed distance dependent on required accuracy.

\subsection{1D lattice Green function: 2D defects}

The introduction of an infinite surface or boundary reduces the
lattice Green function to a one-dimensional column and modifies the
inverse Fourier transformations.  The forces and displacements of
atoms away from a boundary---be it a free surface, grain boundary, or
interface---have a periodicity given by two non-parallel lattice
vectors $\tv_1$ and $\tv_2$ lying in the boundary plane.  The
periodicity is represented in the lattice Green function by the 1D
lattice Green function, $\sum_{mn}
\GL{}(\Rv+m\tv_1+n\tv_2)$.  As with the 3D and 2D lattice Green functions,
evaluation of the 1D lattice Green function is best performed in
Fourier space, and inverse Fourier transforming to real space.  Then,
\be\label{eqn:1dLGF}
\begin{split}
&\CART{G}^{\textrm{L-1D}}(\Rv)
  = \sum_{n_1,n_2=-\infty}^{\infty}
      \CART{G}^{\textrm{L-3D}}(\Rv + n_1\tv_1 + n_2\tv_2) \\
  &= \sum_{n_1,n_2=-\infty}^{\infty}
     \frac{V}{(2\pi)^3} \intBZ d^3k\;
     e^{-i\kv\cdot\Rv} e^{-in_1\kv\cdot\tv_1 -in_2\kv\cdot\tv_2 } \GLk{}(\kv) \\
  &= \sum_{\kplane\in\textrm{BZ}}
     \frac{V}{|\tv_1\x\tv_2|}\intBZone\frac{dk_\perp}{2\pi}\;
     e^{-i(\kperp+\kplane)\cdot\Rv} 
     \GLk{}(\kperp + \kplane)
\end{split}
\ee
where the (finite) summation is over
\beu
\kplane = 2\pi
  \frac{(m_1\tv_1 + m_2\tv_2)\x(\tv_1\x\tv_2)}%
  {|\tv_1\x\tv_2|^2},
\eeu
($m_1$ and $m_2$ integer) that are inside the BZ, and one-dimensional
integration is over $\kperp$ that are parallel to $\tv_1\x\tv_2$ and
inside the BZ.  This is by virtue of the summation over $n_1$ and
$n_2$, similar to the 2D lattice Green function.  \eqn{1dLGF} still
has a pole in $\GLk{}$ to contend with, but it lies purely on the line
where $\kplane = \zero$.  Hence, for $\kplane\ne\zero$, the value of
$\GLk{}=\inv{[\Dk{}]}$ is used, and a discrete inverse Fourier
transform is performed.  Then, the remaining difficulty is the
$1/k_\perp^2$ pole at the gamma point in the 1D inverse Fourier
transform.

The pole at the gamma point in 1D can be split into two contributions
for inverse Fourier transformation: elastic Green function and the
semicontinuum correction.  For one-dimensional variation along
$\kperp$, the elastic Green function is
\beu
\GEk{}(\kperp) = \frac{1}{k_\perp^2}\GEk{},
\eeu
where the factor $\GEk{}=\Le\left(\tv_1\x\tv_2/|\tv_1\x\tv_2|\right)$
depends on $\tv_1\x\tv_2$, and there is no remaining discontinuity at
the gamma point.  Thus, the semicontinuum correction no longer
vanishes at the gamma point, but instead smoothly approaches a
constant value.  Thus, the only piece to be treated analytically is
the $1/k_\perp^{2}$ pole at the origin.

The inverse Fourier transformation of the elastic Green function
requires the evaluation of a single integral.  The elastic Green
function in real space is
\beu
\begin{split}
\GE{}(\Rv) \;[\kpara=\zero]
  &= \frac{V}{|\tv_1\x\tv_2|} \intBZone\frac{dk_\perp}{2\pi}\;
     e^{-i\kperp\cdot\Rv} \GEk{}(\kperp) \fc(k/\kmax) \\
  &= \frac{V}{|\tv_1\x\tv_2|} \GEk{} \Int{-\kmax}{\kmax}\frac{dk}{2\pi}\;
     e^{-ik\Rperp} k^{-2}\fc(k/\kmax),
\end{split}
\eeu
where $\Rperp$ is the (positive) magnitude of $\Rv$ perpendicular to
the plane given by $\tv_1$ and $\tv_2$.  As with the 2D elastic Green
function, the pole is separated off and related to the known Fourier
transform giving
\beu
\begin{split}
& \Int{-\kmax}{\kmax}\frac{dk}{2\pi}\; e^{-ik\Rperp} k^{-2}\fc(k/\kmax)
  = \Int{0}{\kmax} dk\; \frac{\cos(k\Rperp)}{\pi k^2}\fc(k/\kmax) \\
& = \Int{0}{\kmax} dk\; \frac{\cos(k\Rperp)}{\pi k^2} 
  + \Int{\alpha\kmax}{\kmax} dk\; \frac{\cos(k\Rperp)}{\pi k^2}(\fc(k/\kmax)-1) \\
& = \frac{1}{\pi\kmax}\lim_{\delta\to 0}
       \Int{0}{1} du\; \frac{u^2-\delta^2}{(u^2+\delta^2)^2} \cos(u\kmax \Rperp) \\
&\qquad + \Int{\alpha\kmax}{\kmax} dk\; \frac{\cos(k\Rperp)}{\pi k^2}(\fc(k/\kmax)-1) \\
& = -\frac{|\Rperp|}{\pi}\Si(\kmax \Rperp) - \frac{\cos(\kmax \Rperp)}{\pi\kmax} \\
&\qquad  + \Int{\alpha\kmax}{\kmax} dk\; \frac{\cos(k\Rperp)}{\pi k^2}(\fc(k/\kmax)-1).
\end{split}
\eeu
where $\Si(x)\equiv\int_0^x \sin(t)/t\;dt$ is the Sine integral.  The
remaining integral can also be evaluated in closed form, but the
expression is lengthy.  Two important values are $\Rperp=0$ and
$\Rperp\to\infty$, which are
\beu
\begin{split}
\lim_{\Rperp\to0} \GE{}(\Rv) &=
  \frac{V\GEk{}}{|\tv_1\x\tv_2|}
  \frac{3(-1+\alpha^2 - 2\alpha\ln\alpha)}{\pi\kmax(1-\alpha)^3} \\
\lim_{\Rperp\to\infty} \GE{}(\Rv) &=
  -\frac{1}{2}|\Rperp| \frac{V\GEk{}}{|\tv_1\x\tv_2|}.
\end{split}
\eeu
The limiting behavior of $\GE{}\sim |\Rperp|$ from elasticity theory is
recovered.  These results are summarized in \tab{eqn-summary}.

The inverse Fourier transform of the semicontinuum correction
$\Gsck{}$ is performed with a discrete transform on a grid in lines
through the Brillouin zone.  The grid spacing along the line must be
sufficiently small to remove aliasing effects.  As with the 3D and 2D
lattice Green functions, the deviation between the 1D lattice and
elastic Green functions decreases with distance.  Thus, the elastic
Green function may be substituted at a fixed distance, and requiring
the computation of the full lattice Green function for a finite set of
points.

\begingroup
\squeezetable 
\begin{table*}
\caption{Summary of equations for lattice Green function computation
  for different dimensionality.  The split of the lattice Green
  function into three pieces is given for each, along with the angular
  expansion.  The lattice Green function in real space, the limit of
  large $R$, and value at $R=0$ are also given.  The cutoff function
  $\fc$ and parameter $\alpha$ are defined in \eqn{cutoff}; $\kmax$ is
  the radius of a sphere inscribed in the Brillouin zone (BZ).  All
  $\kv$ are restricted to be inside the first BZ.  The finite BZ
  summations are done over a grid of $\Nkpt$ points.  The function
  $\Delta(x)$ is 1 for $x=0$, and 0 elsewhere.  The 3D integrals
  $f^{(0)}_l(x)$, $f^{(2)}_l(x)$ and the 2D integrals $F^{(0)}_n(x)$,
  $F^{(2)}_n(x)$ are defined in \eqn{3dintegrals} and
  \eqn{2dintegrals}, respectively.  For the 2D case, the periodicity
  is defined by a threading lattice vector $\tv$, and $\Rperp$ is the
  magnitude of $\Rv$ perpendicular to $\tv$.  In the 1D case, the
  periodicity is defined by two non-parallel lattice vectors $\tv_1$
  and $\tv_2$; $\Rperp$ is the magnitude of $\Rv$ perpendicular to the
  plane of $\tv_1$ and $\tv_2$.}
\label{tab:eqn-summary}
\beu
\begin{split}
\hline\hline\\[-12pt]
\textrm{3D:\quad}
& \GEk{}(\kv) = \frac{1}{k^2}\sum_{l\textrm{~even}}^{\Lmax}\sum_{m=-l}^l
    \GEk{lm} Y_{lm}(\hat k)
  ,\qquad
  \Gdck{}(\kv) = \sum_{l\textrm{~even}}^{\Lmax}\sum_{m=-l}^l
    \Gdck{lm} Y_{lm}(\hat k)
  ,\qquad
  \Gsck{}(\kv) = \inv{[\Dk{}(\kv)]} - (\GEk{}(\kv) +
    \Gdck{}(\kv))\fc(k/\kmax)
  \\
& \GL{}(\Rv) = \frac{V}{4\pi}\sum_{lm}^{\Lmax} (-1)^{l/2}
    \left[\frac{1}{R}\GEk{lm}f^{(0)}_l(\kmax R)
         +\frac{1}{R^3}\Gdck{lm}f^{(2)}_l(\kmax R) \right] Y_{lm}(\hat R)
    + \frac{1}{\Nkpt}\sum_{\kv\in\textrm{BZ}} e^{-i\kv\cdot\Rv}\Gsck{}(\kv)
  \\
& \GL{}(\Rv\to\infty) = \frac{V}{4\pi}\sum_{lm}^{\Lmax} (-1)^{l/2}
    \left[\frac{1}{R}\GEk{lm} + \frac{l(l+1)}{R^3}\Gdck{lm} \right]
    \left(\prod^{l}_{k\textrm{\ odd}} k \left/ 
          \prod^{l}_{k\textrm{\ even}} k\right.\right) Y_{lm}(\hat R)
    + O(R^{-5})
  \\
& \GL{}(\Rv=0) = \GEk{00} \frac{V\kmax}{2\pi^2}
      \left[1 - \frac{1-\alpha}{2}\right]
    + \Gdck{00} \frac{V\kmax^3}{2\pi^2}
      \left[\frac{1}{3} - \frac{(1-\alpha)(2\alpha^2+5\alpha+8)}{30}\right]
    + \frac{1}{\Nkpt}\sum_{\kv\in\textrm{BZ}} \Gsck{}(\kv)
  \barspace
\textrm{2D:\quad}
& \GEk{}(\kperp) = \frac{1}{k_\perp^2}\sum_{n\textrm{~even}}^{\Nmax}
    \GEk{n} e^{in\phi_k}
  ,\qquad
  \Gdck{}(\kperp) =  \sum_{n\textrm{~even}}^{\Nmax}
    \Gdck{n} e^{in\phi_k}
  ,\qquad
  \Gsck{}(\kv) = \inv{[\Dk{}(\kv)]} - (\GEk{}(\kv) +
    \Gdck{}(\kv))\fc(k/\kmax)\Delta(\kv\cdot\tv\,)
  \\
& \GL{}(\Rv) = \frac{V}{2\pi|\tv\,|}\sum_{n}^{\Nmax} (-1)^{n/2}
    \left[\GEk{n}F^{(0)}_n(\kmax\Rperp)
         +\frac{1}{\Rperp^2}\Gdck{n}F^{(2)}_n(\kmax\Rperp) \right] e^{in\phi_R}
    + \frac{1}{\Nkpt}\sum_{\kpara+\kperp\in\textrm{BZ}} 
      e^{-i(\kpara+\kperp)\cdot\Rv}\Gsck{}(\kpara+\kperp)
  \\
& \GL{}(\Rperp\to\infty) = \frac{V}{2\pi|\tv\,|}\left[
    -\GEk{0}\ln\Rperp + \sum_{n=2}^{\Nmax} (-1)^{n/2}
    \left(\frac{1}{n}\GEk{n} +\frac{n}{\Rperp^2}\Gdck{n}\right) e^{in\phi_R} \right]
    + O(\Rperp^{-4})
  \\
& \GL{}(\Rperp=0) = \GEk{0} \frac{V}{2\pi|\tv\,|}
    \left[ \ln(\kmax) + [\gamma-\ln2]
           + \frac{6\alpha^2(3-\alpha)\ln\alpha
          -(1-\alpha)(5\alpha^2 -22\alpha + 5)}{6(1-\alpha)^3} \right]
  \\
&\qquad\qquad\qquad
    + \Gdck{0} \frac{V\kmax^2}{2\pi|\tv\,|}
      \left[\frac{1}{2} - \frac{(1-\alpha)(3\alpha+7)}{20} \right]
    + \frac{1}{\Nkpt}\sum_{\kpara+\kperp\in\textrm{BZ}} 
      e^{-i\kpara\cdot\Rv}\Gsck{}(\kpara+\kperp)
  \barspace
\textrm{1D:\quad}
& \GEk{}(\kperp) = \frac{1}{k_\perp^2}\GEk{}
  ,\qquad
  \Gdck{}(\kperp) = 0
  ,\qquad
  \Gsck{}(\kv) = \inv{[\Dk{}(\kv)]} - \GEk{}(\kv)\fc(k/\kmax)
    \Delta(\kv\cdot\tv_1)\Delta(\kv\cdot\tv_2)
  \\
& \GL{}(\Rv) =
    \frac{V}{|\tv_1\x\tv_2|} \GEk{}
    \left[ -\frac{|\Rperp|}{\pi}\Si(\kmax\Rperp) - \frac{\cos(\kmax\Rperp)}{\pi\kmax}
       + \Int{\alpha\kmax}{\kmax} dk\; \frac{\cos(k\Rperp)}{\pi k^2}(\fc(k/\kmax)-1)
    \right]
    + \frac{1}{\Nkpt}\sum_{\kpara+\kperp\in\textrm{BZ}} 
      e^{-i(\kpara+\kperp)\cdot\Rv}\Gsck{}(\kpara+\kperp)
  \\
& \GL{}(\Rperp\to\infty) = 
    -\frac{1}{2}|\Rperp| \frac{V\GEk{}}{|\tv_1\x\tv_2|}
    + O(\Rperp^{-1})
  \\
& \GL{}(\Rperp=0) = 
  \frac{V\GEk{}}{|\tv_1\x\tv_2|}
  \frac{3(-1+\alpha^2 - 2\alpha\ln\alpha)}{\pi\kmax(1-\alpha)^3}
    + \frac{1}{\Nkpt}\sum_{\kpara+\kperp\in\textrm{BZ}} 
      e^{-i\kpara\cdot\Rv}\Gsck{}(\kpara+\kperp)
  \\[2pt]
\hline\hline
\end{split}
\eeu
\end{table*}
\endgroup

\section{Error estimation for LGF}
\label{sec:error}

\tab{eqn-summary} shows that as $R$ becomes larger, the
lattice Green function asymptotically matches the elastic Green
function; this matching provides the basis for an error estimate of
the lattice Green function.  The elastic Green function can be
computed knowing only the elastic constants; in turn, the elastic
constants can be computed {\em even for interactions without a fixed
cutoff}, such as \abinit.  Hence, while the dynamical matrix
computational may induce an artificial cutoff, the asymptotic limit of
the lattice Green function is known exactly.  Then an estimate of the
error in the lattice Green function can be determined by estimating
the deviation between the elastic Green function and lattice Green
function.  Surprisingly, an accurate estimate can be obtained using
the elastic constants and the dynamical matrix from an artificially
folded supercell, even if the dynamical matrix has non-zero elements
outside the supercell.  Hence, a single approximate computation of the
dynamical matrix in a supercell together with the elastic constants
provides an estimate of the accuracy of the supercell computation.
This is shown in detail below.

\subsection{Derivation}

The asymptotic connection between the lattice Green function and the
elastic Green function can be understood by viewing the lattice Green
function as a ``numerical grid'' solution to the elastic Green
function differential equation, as in \fig{discrete}.  The mapping of
a continuum differential equation onto a discrete grid with a lattice
equation is a well known method for the numerical solution of
multidimensional partial differential equations.\cite{Richtmyer1967}
The (partial) derivatives can be approximated using finite differences
on the grid.  As the grid spacing becomes small compared to the length
scale of variation of the solution, the continuum solution is
recovered.  Moreover, this mapping can be reversed: given a lattice
equation, taking the limit of zero grid spacing can recover the
continuum partial differential equation.  In the case of the lattice
Green function, the grid is defined by the crystalline lattice, the
lattice equation by \eqn{inverse} and corresponding continuum
differential equation by \eqn{EGF}.

\bfig
\centering
\includegraphics[width=\halfsmallfigwidth]{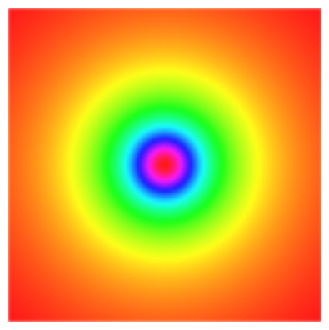}
\includegraphics[width=\halfsmallfigwidth]{grid.eps}

\caption{Connection of solution of continuum differential equation
  mapped onto a lattice equation.  The continuum differential equation
  that defines the solution on the left can be discretized by
  introducing a grid, and approximating derivatives with finite
  differences over the grid to produce a lattice equation.  In the
  limit that the grid spacing becomes small compared to the length
  scale of variation of the solution, the discrete approximation
  matches the continuum solution.  This mapping can also be reversed,
  by starting with a grid and a lattice equation, then taking the
  limit of zero grid spacing to produce a corresponding continuum
  differential equation.}
\label{fig:discrete}
\efig

The analogy of the numerical solution of partial differential
equations provides the basic idea for the estimation of the deviation
between the lattice Green function and elastic Green function.  In
finite difference applications, an estimate of the discretization
error can be determined by substituting the true continuum solution
into the discrete equation, and using Taylor series to approximate the
deviation.\cite{Richtmyer1967} For the lattice Green function
equation, it is the elastic Green function that is an approximation,
but the methodology for error estimation is identical and provides the
deviation between the lattice and elastic Green functions.

The relative deviation of the lattice Green function from the elastic
Green function can be extracted using the real-space lattice Green
function equation \eqn{inverse}.  We begin by defining the relative
deviation $\eGF(\Rv)$, for $R>0$, as
\beu
\GL{}(\Rv) = \GE{}(\Rv)\left\{\ident + \eGF(\Rv)\right\},
\eeu
and substituting into \eqn{inverse} for $R>0$ to get
\beu
\sum_{\xv} \GE{}(\Rv-\xv)\D{}(\xv) +
\sum_{\xv} \GE{}(\Rv-\xv)\eGF(\Rv-\xv)\D{}(\xv) = \zero.
\eeu
Note that while $\GL{}(\Rv)$ is only defined at lattice sites,
$\GE{}(\Rv)$ and $\eGF{}(\Rv)$ are continuum functions.  To simplify
this expression, we introduce the zeroth and second order deviation
functionals $\delZ{}$ and $\del{}$ of a continuum function $f(\Rv)$
around a point $\Rv$,
\beu
\begin{split}
\delZ{f}(\Rv,\xv) \equiv&
  \frac{1}{2}[f(\Rv+\xv)+f(\Rv-\xv)] - f(\Rv), \\
\del{f}(\Rv,\xv) \equiv&
  \frac{1}{2}[f(\Rv+\xv)+f(\Rv-\xv)] - f(\Rv)\\
 &-\frac{1}{2}\xv\cdot\nabla\nabla f(\Rv)\cdot\xv.
\end{split}
\eeu
These functionals describes the deviation between the second-order
finite difference expansion of a function and the true value, with or
without using the Taylor expansion.  For small $x$,
$\delZ{f}(\Rv,\xv)\approx \frac{1}{2} x^2\,df^2/dR^2$ and
$\del{f}(\Rv,\xv)\approx \frac{1}{24} x^4\,df^4/dR^4$.  Since
$\D{}(\xv)=\D{}(-\xv)$, $\delZ{}$ and $\del{}$ can be used to write%
\footnote{It should be noted that the elastic Green function possess a
pole at the origin; hence, the deviation functionals are technically
infinite at $\xv=\pm\Rv$.  This complication is avoided by choosing
finite values for $\protect\GE{}$ and $\protect\eGF{}$ at the origin.
This gives finite values for
$\delZ{\protect\GE{}\protect\eGF}(\Rv,\pm\Rv)$ and
$\del{\protect\GE{}}(\Rv,\pm\Rv)$. While this issue is ignored in the
derivation to avoid complicating the notation further, it does not
affect the results, as the elastic Green function is only evaluated at
lattice sites in our derivation.}
\beu
\begin{split}
& \frac{1}{2}\left[\GE{}(\Rv-\xv)\D{}(\xv)+\GE{}(\Rv+\xv)\D{}(-\xv)\right] \\
&\quad  = \left[ \GE{}(\Rv) + \frac{1}{2} 
      \xv\cdot\nabla\nabla\GE{}(\Rv)\cdot\xv + \del{\GE{}}(\Rv,\xv)
     \right]\D{}(\xv),\\
&\frac{1}{2}\left[\GE{}(\Rv-\xv)\eGF(\Rv-\xv)\D{}(\xv)
                 +\GE{}(\Rv+\xv)\eGF(\Rv+\xv)\D{}(-\xv)\right] \\
&\quad = \left[ \GE{}(\Rv)\eGF(\Rv)
               +\delZ{\GE{}\eGF}(\Rv,\xv) \right]\D{}(\xv).\\
\end{split}
\eeu
Substituting these into the lattice Green function equation gives
\beu
\begin{split}
&\GE{}(\Rv)\left\{\ident + \eGF(\Rv)\right\} \sum_{\xv}\D{}(\xv)
+ \frac{1}{2}\sum_{\xv} \left[
    \xv\cdot\nabla\nabla\GE{}(\Rv)\cdot\xv \right]\D{}(\xv)
\\ &\quad
+ \sum_{\xv}\del{\GE{}}(\Rv,\xv)\D{}(\xv)
+ \sum_{\xv}\delZ{\GE{}\eGF}(\Rv,\xv)\D{}(\xv)
= \zero.
\end{split}
\eeu
The first term is zero because of the dynamical matrix sum rule.  The
second term is simplified to zero by using \eqn{longwaves},
\beu
\begin{split}
&\frac{1}{2}\sum_{\xv}\sum_{abc} x_a x_b \D{cj}(\xv)
  \nabla_a\nabla_b \GE{ic}(\Rv)
\\ &\quad
= -\frac{V}{2}\sum_{abc} (C_{cajb} + C_{cbja})
     \nabla_a\nabla_b \GE{ic}(\Rv) = 0,
\end{split}
\eeu
which is zero by virtue of \eqn{EGF} and applying the
interchangeability of partial derivatives, symmetries of the elastic
Green function and elastic constants: $\GE{ic}=\GE{ci}$,
$C_{cajb}=C_{jbca}=C_{jbac}$, $C_{cbja}=C_{jacb}=C_{jabc}$, and
$\nabla_a\nabla_b=\nabla_b\nabla_a$.  Thus,
\be
\label{eqn:deviation}
\sum_{\xv}\del{\GE{}}(\Rv,\xv)\D{}(\xv)
+ \sum_{\xv}\delZ{\GE{}\eGF}(\Rv,\xv)\D{}(\xv)
= \zero.
\ee

\bfig
\centering
\includegraphics[width=2.5in]{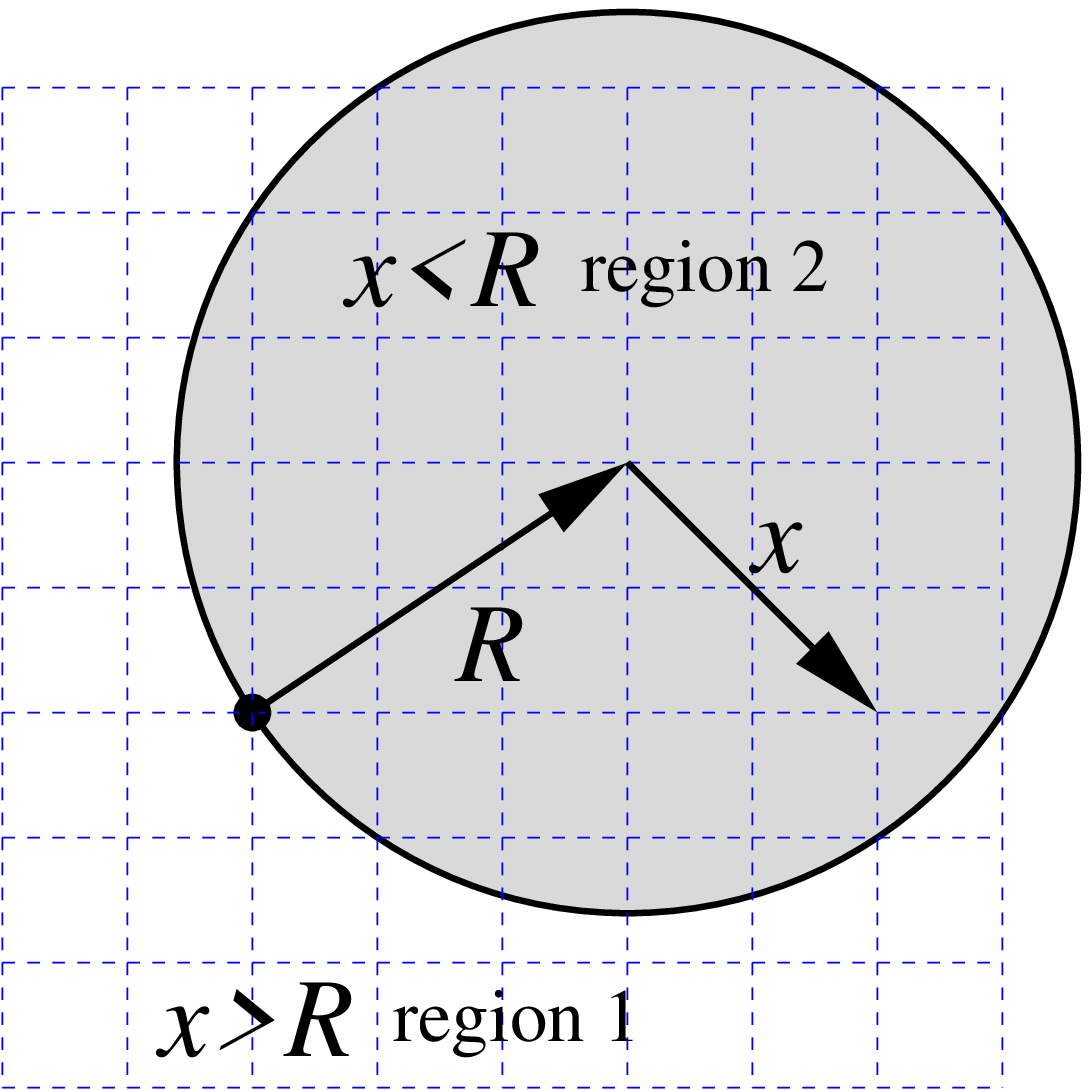}

\caption{Separation of error estimation summation \eqn{deviation} into
  regions 1 and 2.  The different regions can be simplified by using
  different approximations for $\delZ{\GE{}\eGF{}}$ and $\del{\GE{}}$.
  In region 1, $x\gg R$, and the deviations scale as
  $-\GE{}(R)\eGF{}(R)$ and $x^2/R^2\GE{}(R)$, respectively.  For region 2,
  $x\ll R$, and the deviations scale as $x^2/R^2\GE{}(R)\eGF(R)$ and
  $x^4/R^4\GE{}(R)$, respectively.}
\label{fig:regions}
\efig

To evaluate $\eGF(\Rv)$ in \eqn{deviation} requires approximations for
$\delZ{\GE{}\eGF}$ and $\del{\GE{}}$.  These are built by using the
$x\ll R$ and $x\gg R$ asymptotic values over the two regions in
\fig{regions}.  For region 1 ($x\gg R$), both $\GE{}\eGF(\Rv+\xv)$ and
$\GE{}\eGF(\Rv-\xv)$ are much less than $\GE{}(\Rv)\eGF{}(\Rv)$ in
$\delZ{\GE{}\eGF}(\Rv,\xv)$.  For region 2 ($x\ll R$), we use the fact that
$\GE{}(R)\sim 1/R$ and assume that $\eGF(R)\sim 1/R^\beta$ for some
power $\beta$ to be determined.  Using the Taylor series for small $x$
and choosing the maximum $\xdR = 1$ gives the approximation in regions
1 and 2
\beu
\delZ{\GE{}\eGF}(\Rv,\xv) \approx
\GE{}(\Rv)\eGF(\Rv)
\begin{cases}
  -1 & : x>R\\
  \frac{(\beta+1)(\beta+2)}{2} x^2/R^2 & : x\le R.
\end{cases}
\eeu
Similarly, for region 1 ($x\gg R$), $\del{\GE{}}$ is dominated by the
quadratic growth of the last term; as $\GE{}(R)\sim 1/R$, this gives
$x^2/R^2\GE{}(R)$.  Using the next order in the Taylor series for
small $x$ in region 2 ($x\ll R$), with the maximum value of $\xdR=1$
gives the approximation for regions 1 and 2
\beu
\del{\GE{}}(\Rv,\xv) \approx
\GE{}(\Rv)
\begin{cases}
  x^2/R^2 & : x>R\\
  x^4/R^4 & : x\le R.
\end{cases}
\eeu
Substituting these approximations into \eqn{deviation} gives
\beu
\begin{split}
&\GE{}(\Rv)\eGF(\Rv)\left\{
  \frac{(\beta+1)(\beta+2)}{2R^2}\sum_{|\xv|\le R} x^2\D{}(\xv)
  - \sum_{|\xv|>R} \D{}(\xv) \right\}
\\ &\quad
= \GE{}(\Rv)\left\{
  \frac{1}{R^4}\sum_{|\xv|\le R} x^4\D{}(\xv)
  + \frac{1}{R^2}\sum_{|\xv|>R} x^2\D{}(\xv) \right\}.
\end{split}
\eeu
The two sums over $|\xv|>R$ can be simplified by using (1) the sum
rule for the dynamical matrix,
\beu
-\sum_{|\xv|>R}\D{}(\xv) = +\sum_{|\xv|\le R} \D{}(\xv),
\eeu
and (2) using the elastic constants with \eqn{longwaves},
\beu
\begin{split}
\sum_{\xv>|\Rv|}x^2\D{ab}(\xv) &= 
  \sum_{\xv}x^2\D{ab}(\xv) - \sum_{\xv\le|\Rv|}x^2\D{ab}(\xv)
\\ & = -2V\Bigl[ C_{axbx} + C_{ayby} + C_{azbz}
\\ & \qquad -C_{axbx}(R) - C_{ayby}(R) - C_{azbz}(R) \Bigr]
\\ & \equiv 2V\CART{\delta C}_{ab}(R),
\end{split}
\eeu
where $C_{abcd}(R)$ refers to the elastic constants derived using
\eqn{longwaves}, summing only over lattice sites $\xv\le R$.  We will
show shortly that $\eGF(R)\sim R^{-2}$, so after dividing out
$\GE{}(\Rv)$ and substituting $\beta=2$, we have
\be
\label{eqn:deviation-formal}
\begin{split}
&\eGF(\Rv)\left\{
  \frac{6}{R^2}\sum_{|\xv|\le R} x^2\D{}(\xv)
  + \sum_{|\xv|\le R} \D{}(\xv) \right\}
\\ &\quad
= \frac{1}{R^4}\sum_{|\xv|\le R} x^4\D{}(\xv)
  + \frac{2V}{R^2}\CART{\delta C}(R).
\end{split}
\ee

\eqn{deviation-formal} can be further simplified from the small and
large $R$ limits.  In the small $R$ case (large region 1), the first
terms of the left-hand and right-hand sides of \eqn{deviation-formal}
become negligible.  For small $R$, the dominant piece of
$\sum_{|\xv\le R}\D{}(\xv)$ is $\D{}(\zero)$, so
\beu
\eGF_{\textrm{region 1}}(R) \approx 
  \frac{2V}{R^2}\CART{\delta C}(R)\inv{[\D{}(\zero)]}.
\eeu
In the large $R$ case (large region 2), the second terms of the
left-hand and right-hand side of \eqn{deviation-formal} become
negligible.  The summations $\sum_{|\xv|\le R} x^2\D{}(\xv)$ and
$\sum_{|\xv|\le R} x^4\D{}(\xv)$ are related to the $l=0$ spherical
harmonic expansion of $\GEk{}$ and $\Gdck{}$
(c.f.~Appendix~\ref{app:sphereave} and \eqn{sphereave}), so
\beu
\eGF_{\textrm{region 2}}(R) \approx 
  \frac{10}{3R^2}\frac{\Gdck{00}(R)}{\GEk{00}},
\eeu
where $\Gdck{00}(R)$ is evaluated using the truncated dynamical
matrix.  Note that both pieces scale as $R^{-2}$, justifying the
earlier choice of $\beta=2$.  Because the region 1 estimate falls off
faster as $R$ becomes large, and the region 2 estimate goes to zero as
$R$ goes to zero, the final approximation is to sum the two pieces
\be
\label{eqn:deviation-final}
\eGF(R) \approx 
  \frac{2V}{R^2}\CART{\delta C}(R)\inv{[\D{}(\zero)]}
 + \frac{10}{3R^2}\frac{\Gdck{00}(R)}{\GEk{00}}.
\ee

The main feature of \eqn{deviation-final} is that the two region
estimates can be determined using a {\em single} supercell
calculation, even if the dynamical matrix lacks a finite interaction
cutoff.  The region 1 estimate is computed by comparing the true
elastic constants to the elastic constants computed from a folded
dynamical matrix, and using the supercell dimension for $R$.  The
dynamical matrix can result from a direct force computation in a
finite supercell, or evaluating $\Dk{}$ for a finite k-point grid.  In
the latter case, the inverse grid spacing provides the value for $R$.
The region 2 estimate is computable because $\Gdc{}$ is only summed
over $\xv\le R$ for $\D{}(\xv)$.  In both of these cases, it is
assumed that the effect of folding the dynamical matrix into the
supercell is approximately equivalent to truncating it outside the
supercell.

\subsection{Numerical example of error estimate: simple-cubic lattice}

While \eqn{deviation-final} has the advantage of being computable for
long-ranged dynamical matrices, it is not clear if too much accuracy
has been lost in the series of approximations; so a numerical example
is used to highlight the range of applicability.  A series of
pseudo-random long-range dynamical matrices are generated on a
simple-cubic lattice with characteristics related to real material
systems; and for each, the lattice Green function, relative deviation
to the elastic Green function, and region 1 and 2 estimates are
computed.

The dynamical matrices are generated using $\D{}(R)\sim \sin(\pi
R/a_0) R^{-4}$, cutoff at $25a_0$, with lattice constant $a_0=1$.  The
functional form is chosen to provide a long-range interaction, whose
falloff is still fast enough to produce finite elastic constants in
\eqn{longwaves}.  The $\sin(\pi R/a_0)$ functional form produces a
Friedel-like oscillation, as might be expected in a metallic system.
The dynamical matrix elements at each site are pseudo-random numbers
from a Gaussian distribution with mean 0 and standard deviation
$\sin(\pi R/a_0)R^{-4}$.  The dynamical matrix is symmetrized using
the cubic point group.  The elastic constants and phonons are
computed; if there are unstable phonons, or the elastic anisotropy is
greater than 3, the dynamical matrix is rejected.  100 random, stable,
long-range simple-cubic dynamical matrices are generated in this
manner; for each, the lattice and elastic Green functions along with
the relative deviation are generated.  The dynamical matrix is
``folded down'' into supercells from $2\x2\x2$ to $20\x20\x20$ to
compute the region 1 and 2 estimates in \eqn{deviation-final}.

\bfig
\centering
\includegraphics[width=\smallfigwidth]{sc.err.rel.combined.eps}

\caption{Relative deviation between EGF and LGF for simple-cubic test
  case.  The points give the deviation between the lattice Green
  function computed with the full dynamical matrix and the elastic
  Green function.  The region 1 and 2 estimates are computed using the
  folded dynamical matrix in cubic supercells, and combined as in
  \eqn{deviation-final} to produce the total deviation estimate.  (a)
  Single random dynamical matrix shows an individual example of error
  estimation.  (b) 100 different random dynamical matrices were
  computed, along with their associated LGF's.  The average deviation
  over the ensemble average shows that we have an accurate computation
  of the error, even for the case of small supercells ($2\x2\x2$) with
  the long-range dynamical matrix (cutoff at $25a_0$).}
\label{fig:sc}
\efig

\fig{sc} shows the true deviation and estimates from our test case for
both a single example, and the average results from the 100 dynamical
matrices.  As expected from the derivation, the region 1 estimate
dominates for small $R$, and falls off as the supercell becomes large
enough to accurately produce the elastic constants.  The region 2
estimate becomes important for large $R$, capturing the long-range
effect from the discontinuity correction.  What is especially
encouraging is that the error estimate is accurate even for small
supercells---such as $2\x2\x2$---where the supercell dynamical matrix
calculation is clearly inaccurate due to the long range.  This is
perhaps the most impressive feature of \eqn{deviation-final}: Even
when the dynamical matrix calculation comes from a small supercell,
the known elastic constants can still provide an accurate error
estimate {\em without} requiring comparisons to larger supercells.
Hence, a supercell-size effect estimate on the lattice Green function
computation is provided from a {\em single} supercell dynamical matrix
computation.

\section{Discussion}
\label{sec:concl}

The deviation between the lattice Green function and elastic Green
function in \eqn{deviation-final} can be described by a single length
scale $\lelas$ that characterizes the recovery of continuum elastic
behavior from atomistic lattice behavior: $\eGF(R)\approx
(\lelas/R)^2$.  This length scale determines the range out to which
the lattice Green function should be computed in lieu of the elastic
Green function.  For example, if the magnitude of the largest lattice
vector $R_{\textrm{max}}$ is greater than $10\lelas$, the lattice
Green function can be computed for lattice vectors $|\Rv|<10\lelas$,
and the elastic Green function used for the remainder, while
introducing a total error of $1\%$.  This choice can greatly speed the
computation of the lattice Green function for large simulations by (1)
limiting the $k$-point grid size, and (2) restricting the set of
points over which the full lattice Green function must be computed.

The length scale $\lelas$ is also a fundamental length scale for
quasi-continuum\cite{Rudd2005} and flexible boundary conditions
methods\cite{Thomson1992,Rao1998} where it determines the range at
which the relaxation response using elastic finite-elements or the
bulk continuum is accurate compared to atomistic response.  This
length is not necessarily the same as the interaction force
cutoff---it may be larger or smaller.  For example, the region 2
estimate of deviation for an isotropic nearest-neighbor interaction
gives $\lelas = \sqrt{1/6}R_{\textrm{nn}} \approx 0.4
R_{\textrm{nn}}$, which suggests transitioning from atomistic to
finite-elements at twice the interaction cutoff produces errors on the
order of $4\%$ in position.  On the other extreme, \abinit\ 
calculations in metals have shown surprisingly small $\lelas$,
considering the known long-range interactions in metallic
systems.\cite{Woodward2001} It is the small value of $\lelas$ that has
allowed the accurate calculation of isolated dislocations using
flexible boundary condition methods in \abinit.
Knowledge of $\lelas$ is essential to constructing accurate
computational cells that are large enough to produce accurate
response, but do not waste computational resources treating
interactions that can be replaced with elastic response.

This paper presents an accurate computational algorithm for the
lattice Green function from limited dynamical matrix information
together with the elastic constants.  In conjunction, an accurate
error estimate using the limited dynamical matrix computable from a
{\em single} supercell computation allows measurement of the
supercell-size effect.  The error estimate produces a length scale
$\lelas$ which characterizes the crossover from atomistic harmonic
response to continuum elastic response.  The algorithm for lattice
Green function computation together with the determination of
crossover length scale has already allowed the accurate computation of
single extended dislocation defects using
\abinit.\cite{Woodward2001,Woodward2002,Woodward2004,TrinkleSSS05} The
approach can also be utilized to implement flexible boundary condition
methods for point defects, crack opening and tip propagation, surfaces
and boundaries coupled with \abinit: providing chemically accurate
interactions coupled with correct treatment of the long-range elastic
response of extended defects.

\begin{acknowledgments}
The author thanks R.~Hennig, S.~Rao, J.~Wilkins, and C.~Woodward for
helpful discussions.  This research was performed while DRT held a
National Research Council Research Associateship Award at AFRL.
Computational resources were provided by the Ohio Supercomputing
Center.
\end{acknowledgments}

\appendix

\section{Angular integration in inverse Fourier transform }
\label{app:angular}

The integration of the angular portion of the inverse Fourier
transform in three dimensions can be performed analytically.
Particular integrals referenced below can be found in \INT.

Evaluation of the three dimensional integral
\be\label{eqn:3dangular}
  \iint_{4\pi}\displaylimits \negthickspace d^2\hat k\; 
     e^{ikR(\hat k\cdot\hat R)} Y_{lm}(\hat k),
\ee
begins by rotating the variable of integration to a new coordinate
system given by $\hat p(\hat k)$ such that $\hat R$ aligned along the
$p_z$-axis.  The spherical harmonic $Y_{lm}(\hat k)$ is written as a
linear expansion of $Y_{lm'}(\hat p)$
\be\label{eqn:Ylm_exp}
Y_{lm}(\hat k) = \sum_{m'=-l}^l a^{(l)}_{mm'}(\hat R) Y_{lm'}(\hat p),
\ee
where the $a^{(l)}_{mm'}(\hat R)$ coefficients will be determined
later.  With this expansion, \eqn{3dangular} becomes
\beu
\sum_{m'=-l}^l  a^{(l)}_{mm'}(\hat R)
  \int_0^\pi d\theta_p\; \sin\theta_p e^{ikR\cos\theta_p}
  \int_0^{2\pi} d\phi_p\; Y_{lm'}(\theta_p,\phi_p),
\eeu
where $(\theta_p,\phi_p)$ are the angular coordinates of $\hat p$ with
$\Rv$ as the $z$-axis.  In this coordinate system, $\theta_p$ is the
angle between $\hat p$ and $\hat R$; hence, $\cos\theta_p = \hat
k\cdot \hat R$.  The integral simplifies by (1) transforming
$u=\cos\theta_p$, and (2) noting that the $\phi_p$ integral is
non-zero only for $m'=0$.  Thus, our integral reduces to
\beu
a^{(l)}_{m,0}(\hat R) \sqrt{\pi}\sqrt{2l+1}
  \int_{-1}^1 du\; P_l(u) e^{ikRu}.
\eeu
The integral of the Legendre polynomial are expressions 7.393.1 and
7.393.2 in \INT, which has
\beu
\int_{-1}^1 du\; P_l(u) e^{ixu} = 2(i)^l j_l(x),
\eeu
where $j_l(x)$ is the regular spherical Bessel
function,\cite{MathFunctions} $\sqrt{\pi/2x}\,J_{l+1/2}(x)$.

Finally, $a^{(l)}_{m,0}(\hat R)$ must be evaluated to produce the
final expression.  \eqn{Ylm_exp} can multiplied by $Y^*_{ln}(\hat p)$
and integrated over $4\pi$ to give
\beu
\begin{split}
a^{(l)}_{mn}(\hat R) 
  &= \intsphere d^2\hat p\; Y_{lm}(\hat k(\hat p)) Y^*_{ln}(\hat p)\\
  &= \intsphere d^2\hat k\; Y_{lm}(\hat k) Y^*_{ln}(\hat p(\hat k)).
\end{split}
\eeu
Then, the $n=0$ component is
\beu
a^{(l)}_{m,0}(\hat R) 
  = \intsphere d^2\hat k\; Y_{lm}(\hat k) \sqrt{\frac{2l+1}{4\pi}} P_{l}(\cos\theta_p),
\eeu
where $\theta_p$ is the angle between $\hat p(\hat k)$ and the $z$-axis
in $p$'s coordinate system.  Given our rotation, $\cos\theta =
\hat k\cdot\hat R$.  The addition theorem for spherical harmonics,
\beu
P_l(\hat k\cdot\hat R) = \frac{4\pi}{2l+1}
  \sum_{m'=-l}^l Y_{lm'}(\hat R) Y^*_{lm'}(\hat k),
\eeu
then gives
\beu
\begin{split}
a^{(l)}_{m,0}(\hat R) 
  & = \sqrt{\frac{4\pi}{2l+1}} \sum_{m'=-l}^l Y_{lm'}(\hat R)
    \intsphere d^2\hat k\; Y_{lm}(\hat k) Y^*_{lm'}(\hat k) \\
  & =  \sqrt{\frac{4\pi}{2l+1}} Y_{lm}(\hat R).
\end{split}
\eeu
Combining the terms gives
\be\label{eqn:3dangularfinal}
  \intsphere d^2\hat k\; 
     e^{ikR(\hat k\cdot\hat R)} Y_{lm}(\hat k) = 4\pi (i)^l j_l(kR)
Y_{lm}(\hat R).
\ee

\section{Region 2 error estimate}
\label{app:sphereave}

The region 2 error estimate in \eqn{deviation-formal} contains two
summations over $\xv$ with $\D{}(\xv)$ which can approximated using
the $l=0$ spherical harmonic components of $\GEk{}$ and $\Gdck{}$.
The quartic term $\sum_{\xv} x^4\D{}(\xv)$ appears in the computation
of $\Gdck{00}$:
\beu
\Gdck{00}=\frac{1}{\sqrt{4\pi}} \intsphere d^2\hat k\;
  \GEk{}(\hat k)\left[-\frac{1}{24}
    \sum_{\xv}(\xv\cdot\hat k)^4\D{}(\xv) \right]
  \GEk{}(\hat k).
\eeu
We can approximate $\GEk{}(\hat k)$ with its spherical average value,
$\GEk{00}/\sqrt{4\pi}$.  The integral $\intsphere d^2\hat
k\;(\xv\cdot\hat k)^4 = 4\pi x^4/5$, so
\beu
\left[-\sum_{\xv} x^4\D{}(\xv) \right] \approx
  120\sqrt{4\pi} \inv{[\GEk{00}]} \Gdck{00} \inv{[\GEk{00}]}.
\eeu
The $\sum_{\xv} x^2\D{}(\xv)$ term requires more egregious
approximations.  Starting with the definition of $\GEk{00}$,
\beu
\GEk{00}=\frac{1}{\sqrt{4\pi}} \intsphere d^2\hat k\;
  \inv{\left[-\frac{1}{2}\sum_{\xv} (\xv\cdot\hat k)^2 \D{}(\xv)
  \right]},
\eeu
both sides are inverted, and the inverse of the spherical average is
approximated as the average of the inverse,
\beu
\begin{split}
\inv{[\GEk{00}]}\sqrt{4\pi} &\approx \frac{1}{4\pi}\intsphere d^2\hat k\;
  \left[-\frac{1}{2}\sum_{\xv}\D{}(\xv)
                      (\xv\cdot\hat k)^2\right]
\\ &=
  \frac{1}{4\pi}\frac{4\pi}{3}\frac{1}{2} \left[-\sum_{\xv}x^2\D{}(\xv)\right].
\end{split}
\eeu
Inverting again gives
\beu
\inv{\left[-\sum_{\xv} x^2\D{}(\xv) \right]} \approx
  \frac{1}{6\sqrt{4\pi}} \GEk{00}.
\eeu
Then, for the large $R$ limit of \eqn{deviation-formal}, we have
\be
\label{eqn:sphereave}
\frac{1}{6R^2}\inv{\left[-\sum_{\xv} x^2\D{}(\xv) \right]}
\left[-\sum_{\xv} x^4\D{}(\xv) \right]
\approx
\frac{10}{3R^2}\frac{\Gdck{00}}{\GEk{00}}.
\ee
The final approximation is to limit the summations to $|\xv|\le R$; for
that evaluation, we replace the true $\Gdck{00}$ with $\Gdck{00}(R)$,
evaluated using summations over $|\xv|\le R$.

Despite the approximations at use in \eqn{sphereave}, it is exact in
an important limit: elastically isotropic crystals.  Since the elastic
Green function is isotropic, $\GEk{}(\hat k) = \GEk{00}/\sqrt{4\pi}$,
and the approximations in inverting $\GEk{00}$ are exact.
Furthermore, for nearest-neighbor interactions, the left-hand side of
\eqn{sphereave} is exactly $R_{\textrm{nn}}^2/6R^2$, making $\lelas =
\sqrt{1/6}R_{\textrm{nn}}$.


\end{document}